\documentclass[structabstract]{aa}  
\usepackage{graphicx,txfonts,lscape,natbib}
\usepackage{txfonts}
\bibpunct{(}{)}{;}{a}{}{,} 
\begin{document}

   \title{Search for free-floating planetary-mass objects in the Pleiades}


   \author{M$.$ R$.$ Zapatero Osorio\inst{1}
          \and
                M$.$ C$.$ G\'alvez Ortiz\inst{1}
          \and
                G$.$ Bihain\inst{2}
          \and
                C$.$ A$.$ L$.$ Bailer-Jones\inst{3}
          \and
                R$.$ Rebolo\inst{4,5}
          \and
                Th$.$ Henning\inst{3}
          \and
                S$.$ Boudreault\inst{6}
          \and
                V$.$ J$.$ S$.$ B\'ejar\inst{4,5}
          \and
                B$.$ Goldman\inst{3}
          \and
                R$.$ Mundt\inst{3}
          \and
                J$.$ A$.$ Caballero\inst{7}
          }

   \institute{Centro de Astrobiolog\'\i a (CSIC-INTA), Carretera de Ajalvir km 4, E-28850 Torrej\'on de Ardoz, Madrid, Spain\\
              \email{mosorio,mcz@cab.inta-csic.es}
         \and
               Leibniz-Institut f\"ur Astrophysik Potsdam (AIP), An der Sternwarte 16, 14482, Potsdam, Germany\\
             \email{gbihain@aip.de}
         \and
               Max-Planck-Institut f\"ur Astronomie, K\"onigstuhl 17, 69117 Heidelberg, Germany\\
              \email{calj,henning,goldman,mundt@mpia.de}
         \and
               Instituto de Astrof\'\i sica de Canarias, V\'\i a L\'actea s/n, E-38205 La Laguna, Tenerife, Spain\\
              \email{rrl@iac.es}
         \and
               Dept$.$ Astrof\'\i sica, Universidad de La Laguna, E-38206 La Laguna, Tenerife, Spain\\ 
              \email{vbejar@iac.es}
         \and
               GEPI, Observatoire de Paris, CNRS, Universit\'e Paris Diderot; 5 Place Jules Janssen, F-92190 Meudon, France\\
              \email{Steve.Boudreault@obspm.fr}
         \and
               Centro de Astrobiolog\'\i a (CSIC-INTA), PO Box 78, E-28691 Villanueva de la Ca\~nada, Madrid, Spain\\
              \email{caballero@cab.inta-csic.es}
             }

   \date{Received ; accepted }
 
  \abstract
  {} 
   {We aim at identifying the least massive population of the solar metallicity, young (120 Myr), nearby (133.5 pc) Pleiades star cluster with the ultimate goal of understanding the physical properties of intermediate-age, free-floating, low-mass brown dwarfs and giant planetary-mass objects, and deriving the cluster substellar mass function across the deuterium-burning mass limit at $\approx$\,0.012 M$_\odot$.}
   {We performed a deep photometric and astrometric $J$- and $H$-band survey covering an area of $\sim$0.8 deg$^2$ in the Pleiades cluster. The images with completeness and limiting magnitudes of $J,H \approx 20.2$ and $\approx 21.5$ mag were acquired $\sim$\,9 yr apart, allowing us to derive proper motions with a typical precision of $\pm$6 mas\,yr$^{-1}$. For the cluster distance and age, the survey is sensitive to Pleiades members with masses in the interval $\approx$0.2--0.008 M$_\odot$. $J$- and $H$-band data were complemented with $Z$, $K$, and mid-infrared magnitudes up to 4.6 $\mu$m coming from the UKIRT Infrared Deep Sky Survey (UKIDSS), the {\sl WISE} catalog, and follow-up observations of our own. Pleiades member candidates were selected to have proper motions compatible with that of the cluster, and colors following the known Pleiades sequence in the interval $J$ = 15.5--18.8 mag, and $Z_{\rm{UKIDSS}} - J \ge 2.3$ mag or $Z$ nondetections for $J > 18.8$ mag. 
 }
   {We found a neat sequence of astrometric and photometric Pleiades substellar member candidates with two or more proper motion measurements and with magnitudes and masses in the intervals $J$ = 15.5--21.2 mag and $\approx$0.072--0.008 M$_\odot$. The faintest objects show very red near- and mid-infrared colors exceeding those of field high-gravity dwarfs by $\ge$0.5 mag. This agrees with the reported properties of field young L-type dwarfs and giant planets orbiting stars of ages  of $\sim$100 Myr. The Pleiades photometric sequence does not show any color turn-over because of the presence of photospheric methane absorption down to $J$ = 20.3 mag, which is about 1 mag fainter than predicted by the combination of evolutionary models and colors computed from model atmospheres. The astrometric data suggest that Pleiades brown dwarfs have a proper motion dispersion of 6.4--7.5 mas\,yr$^{-1}$, and are dynamically relaxed at the age of the cluster. The Pleiades mass function extends down to the deuterium burning-mass threshold, with a slope fairly similar to that of other young star clusters and stellar associations. The new discoveries may become benchmark objects for interpreting the observations of the emerging young ultracool population and giant planets around stars in the solar neighborhood.    }
   {}

   \keywords{stars: low-mass, brown dwarfs, late type -- open clusters and associations: individual: the Pleiades}

   \maketitle
%

\section{Introduction}
Understanding the evolution and physical properties of substellar objects is one major question in modern Astrophysics. To achieve this goal, it is mandatory to identify sources at various ages to construct a picture of the complete evolutionary sequence. At the youngest ages of fewer than 10 Myr, various groups have deeply investigated different star-forming regions and star clusters in Orion, Upper Scorpio, Taurus, and other nearby young stellar associations (e.g., \citealt{lucas00,luhman08,luhman09,lodieu12,lodieu13,palau12,pena12}; see review by \citealt{luhman12}). The smallest objects found have masses of about 0.004--0.006 M$_\odot$ and spectral types around mid-L. T dwarf candidates have also been discovered in the direction of very young star clusters and moving groups (e.g., \citealt{osorio02,burgess09,bihain09,pena11,lodieu11,delorme12,spezzi12,parker13,naud14}), although their nature and cluster membership are still debated. In the field, all-sky surveys such as the DEep Near-Infrared Survey (DENIS, \citealt{epchtein97}), the Two Micron All Sky Survey (2MASS, \citealt{skrutskie06}), and the Wide-field Infrared Survey Explorer ({\sl WISE}, \citealt{wright10}) combined with other large-scale surveys such as the Sloan Digital Sky Survey \citep{gunn95} and the Infrared Deep Sky Survey (UKIDSS, \citealt{lawrence07}), have led to exciting discoveries of L-, T-, and Y-type dwarfs (e.g., \citealt{delfosse97,tinney98,strauss99,kirk00,kirk11,kirk13,delorme10,cushing11,burningham13}). The great majority of these findings most likely have an age $\ge$1 Gyr, consistent with a field population of brown dwarfs in the solar neighborhood. Young field brown dwarfs and planetary-mass objects of similar spectral classification (the latter have a mass below the deuterium burning-mass threshold at $\approx$0.012 M$_{\odot}$, \citealt{saumon96}) are rarer, but are found at a rate of 8--20\%~(e.g., \citealt[and references therein]{cruz09}). Because young substellar objects with less than a few to several hundred Myr are expected to undergo gravitational self-contraction, their atmospheres have low pressures (or low surface gravities)
and consequently display spectroscopic features that distinguish them from their older, high-gravity counterparts (e.g., \citealt{martin10,allers13}): enhanced TiO and VO bands, narrow or weak alkali lines, and a triangular $H$-band spectral shape. Unfortunately, their ages and metallicities are poorly constrained, and trigonometric parallaxes are lacking for most of them.

Being one of the nearest young open clusters to the Sun, the Pleiades offers an unique opportunity to scrutinize brown dwarfs and planetary-mass objects of known (intermediate) age and metallicity. With a solar abundance \citep{soderblom09}, an age of 120 Myr \citep{basri96,martin98}, and a distance of 133.5 pc \citep{soderblom05}, the Pleiades is a rich cluster containing over 1000 members \citep{sarro14}. The latest studies on the Pleiades substellar regime have probed cluster members with masses of $\sim$0.020--0.025 M$_\odot$ and early-L spectral types \citep{bihain06,bihain10,lodieu12}. The first attempt to identify Pleiades T dwarfs was made by \citet{casewell07,casewell10,casewell11}. The high proper motion of the cluster [(19.71, $-$44.82) mas\,yr$^{-1}$, \citealt{lotkin03}] is distinctive from the field \citep{hambly93}, making astrometric selection of member candidates straightforward. Studying Pleiades brown dwarfs and planetary-mass objects helps to constrain and improve substellar evolutionary models. Furthermore, Pleiades members may become crucial for understanding objects of similar temperatures in the field as well as planets that
orbit stars.

Here, we report on a deep photometric and astrometric exploration using broad-band filters from 1 through 5 $\mu$m and covering an area of 0.8 deg$^2$ to identify Pleiades substellar member candidates with masses across the deuterium burning-mass limit. This survey is about 1--2 mag deeper in the $JH$-bands than any previous work carried out in the Pleiades so far. According to current evolutionary models \citep{chabrier00a}, Pleiades members in the mass range from the substellar borderline to about 0.01 M$_\odot$ would have effective temperatures spanning from 2750 to 900 K. We define the Pleiades sequence at very low masses and compare the photometric properties of our candidates with those of field dwarfs of similar temperatures (late-M, L, and T spectral types) and planets orbiting stars known in the literature. The 120 Myr theoretical isochrone is tested against our observations. Finally, we derive that brown dwarfs are dynamically relaxed in the cluster, and we construct the Pleiades substellar mass function, proving it to be essentially similar to that of younger star clusters.


\section{Observations and data analysis}

\subsection{Near-infrared images \label{observations}}
\begin{figure}
   \centering
   \includegraphics[width=0.48\textwidth]{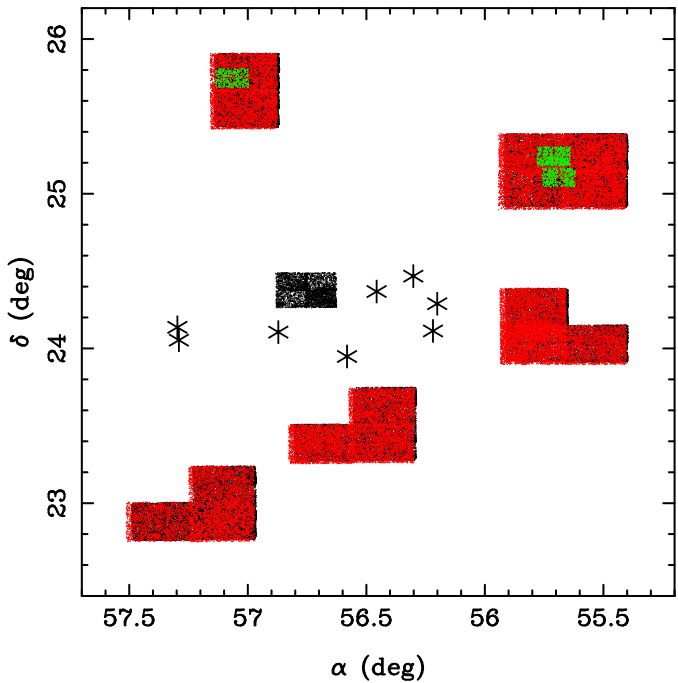}
   \caption{Layout of the $JH$ survey in the ($\alpha, \delta$) plane (equinox J2000). First-epoch $J$-band data are plotted in black, second-epoch $H$-band detections are shown in red. Green stands for the HAWK-I methane images. For reference, eight well-known bright stars of the Pleiades are indicated with asterisks. The common $JH$ area is $\sim$0.8 deg$^2$. North is up and east is to the left.}
              \label{survey}
    \end{figure}

Deep $J$- (1.2 $\mu$m) and $H$-band (1.65 $\mu$m) images were acquired with the prime focus, wide-field infrared cameras Omega-Prime and Omega-2000 of the 3.5-m telescope on the Calar Alto Observatory in Almer\'\i a (Spain). Omega-Prime has a Rockwell 1024\,$\times$\,1024 pixel HgCdTe Hawaii array (1--2.5 $\mu$m) with a pixel pitch of 0\farcs3961 on the sky, giving a field of view of 6\farcm76\,$\times$\,6\farcm76. Omega-2000 delivers a larger field of view of 15\farcm36\,$\times$\,15\farcm36 provided by its 2048\,$\times$\,2048 Hawaii-2 detector with a pixel scale of 0\farcs45 on the sky. The first set of imaging data were obtained with the $J$-band and Omega-Prime between October 29 and 31 in 1998. The total area covered in the Pleiades cluster was $\sim$0.9 deg$^2$ (60 Omega-Prime pointings). Nine years later, we acquired the second set of images using the $H$-band filter and Omega-2000 between October 29 and December 1 in 2007. The area covered by Omega-2000 entirely overlaps with that of Omega-Prime, except for a small portion immediately to the north of the cluster center. The common $J$- and $H$-band surveyed region is $\sim$0.8 deg$^2$, covering patchy areas to the north, west, and south of the accepted Pleiades cluster center, as shown in Figure~\ref{survey}, and avoiding very bright stars that would cause strong saturation of the detectors and the Merope nebula, which is a source of significant extinction \cite[and references therein]{stauffer87}. All $JH$ data were taken under photometric conditions with stable seeing between 0\farcs9 and 1\farcs2. Total on-source exposure times ranged from 9.6 to 14.9 min for the $J$-band data, and from 1.5 to 2.55 h for the $H$-band images, depending on the seeing. These $JH$ data (main survey) represent the core of our photometric and astrometric search for the Pleiades least-massive population.

Additional supporting near-infrared images were acquired with the narrow-band methane filter and the High-Acuity, Wide-field $K$-band Imaging (HAWK-I) instrument \citep{pirard04} installed at one of the Nasmyth foci of the fourth Very Large Telescope (VLT) on Paranal Observatory (Chile), and with the $J$- and $K_s$-bands and the Long-slit Intermediate Resolution Infrared Spectrograph (LIRIS, \citealt{manchado98}) attached to the Cassegrain focus of the 4.2-m William Herschel Telescope (WHT) on Roque de los Muchachos Observatory (La Palma, Spain). HAWK-I delivers a on-sky field of view of 7\farcm5\,$\times$\,7\farcm5 with a cross-shaped gap of 15\arcsec~between the four Hawaii 2RG 2048\,$\times$\,2048 pixel detectors. The pixel scale is 0\farcs106. The central wavelength of the methane filter is 1.575 $\mu$m, meaning that it explores wavelengths blueward of the strong methane absorption occurring at 1.6 $\mu$m, and is well suited for studying T dwarfs (e.g., \citealt{pena11}). LIRIS uses a 1024\,$\times$\,1024 Hawaii detector with a pixel size of 0\farcs25 on the sky, yielding a field of view of 4\farcm27\,$\times$\,4\farcm27. HAWK-I images were acquired as part of program 088.C-0328 in queue mode between 2011 October 21 and December 30, and LIRIS data were taken in service and visitor modes on 2013 September 20 and October 10--13. Both HAWK-I and LIRIS observations were intended to follow-up a few candidates by obtaining additional photometry and proper motion measurements, therefore they cover much smaller areas ($\sim$0.04 deg$^2$) than the large $JH$ main survey. Observing conditions were photometric, and seeing oscillated between 0\farcs5 and 1\farcs2. Exposure times were 24 min and 56 min for HAWK-I, and ranged from 18 to 66 min for LIRIS. 

All images were observed employing a similar strategy, which included sequences of short individual exposures ($\le$80 s, depending on filter, seeing, and airmass during the observations) following a multipoint dither pattern for a proper subtraction of the sky background contribution. Raw data were processed using standard techniques and routines within the IRAF\footnote{IRAF is distributed by National Optical Astronomy Observatories, which is operated by the Association of Universities for Research in Astronomy, Inc., under contract with the National Science Foundation.} (Image Reduction and Analysis Facility) environment. Sky-subtracted frames were divided by their corresponding flat-field images, registered, and stacked to produce deep data. 

Final $J$ and $H$ frames of the main survey were calibrated in right ascension ($\alpha$) and declination ($\delta$) using $\alpha$ and $\delta$ coordinates (equinox J2000) of the Galactic Clusters Survey (GCS) program included in UKIDSS. Only common sources with tabulated photometric error bars lower than $\pm$0.1 mag were used for the astrometric calibration. The precision achieved was typically of a few tenths of a pixel ($\pm$0\farcs1--0\farcs3).

\subsection{Photometry \label{phot}}
Aperture and point-spread-function (PSF) photometry was carried out using routines within the IRAF DAOPHOT package. First, sources were automatically identified with an algorithm that distinguishes between point-like and well-resolved objects (clear galaxies). Resolved sources with a full width at half-maximum (FWHM) slightly higher than the mean seeing were not removed from the final catalogs of detected objects, particularly at the faintest magnitudes, where the uncertainties associated with FWHM measurements are large. For each frame, we found the best-fit PSF by selecting $\ge$5 isolated reference stars that were homogeneously distributed across the field. Instrumental $JHK_s$ magnitudes were converted into observed magnitudes using the UKIDSS GCS photometry of sources with cataloged error bars better than $\pm$0.1 mag. Typically, we were able to find over 30 of these UKIDSS sources in common with Omega-Prime, Omega-2000, and LIRIS separated frames. The internal UKIDSS photometric calibration is addressed by \citet{hewett06} and \citet{hodgkin09}. The typical error of the photometric calibration is $\pm$0.04 mag for all filters in our study. 

The $JH$ main survey extends from $J,H \sim 13$ mag (excluding the bright nonlinear regime of the detectors) down to $\sim$21.5 mag at a 3--4 $\sigma$ detection level, where $\sigma$ stands for the sky background noise. We remark that not all images are as deep, particularly among the first-epoch $J$ data; the most shallow images roughly reach $\sim$21 mag. Completeness (defined at 10 $\sigma$ above the sky background noise) is about $J, H = 20.2$ mag. This survey is about 2 mag deeper than the observations of the Pleiades cluster carried out as part of the UKIDSS GCS project (see \citealt{lodieu12}). According to the evolutionary models provided by \citet[and references therein]{chabrier00a}, at the age, metallicity, and distance of the Pleiades, our main survey is sensitive to all masses from low-mass stars of $\sim$0.2 M$_\odot$ to isolated planetary-mass objects of $\sim$0.008 M$_\odot$ (or $\sim$8 M$_{\rm Jup}$, corresponding to the limiting magnitude). The survey thus covers a factor of 25 in mass passing through two substellar boundaries: the hydrogen-burning mass limit at 0.072 M$_\odot$ (\citealt{chabrier97}; star--brown dwarf separation) and the deuterium-burning mass threshold at 0.012 M$_\odot$ (\citealt{saumon96,burrows97,chabrier00b}; planetary frontier).

   \begin{figure}
   \centering
   \includegraphics[width=0.48\textwidth]{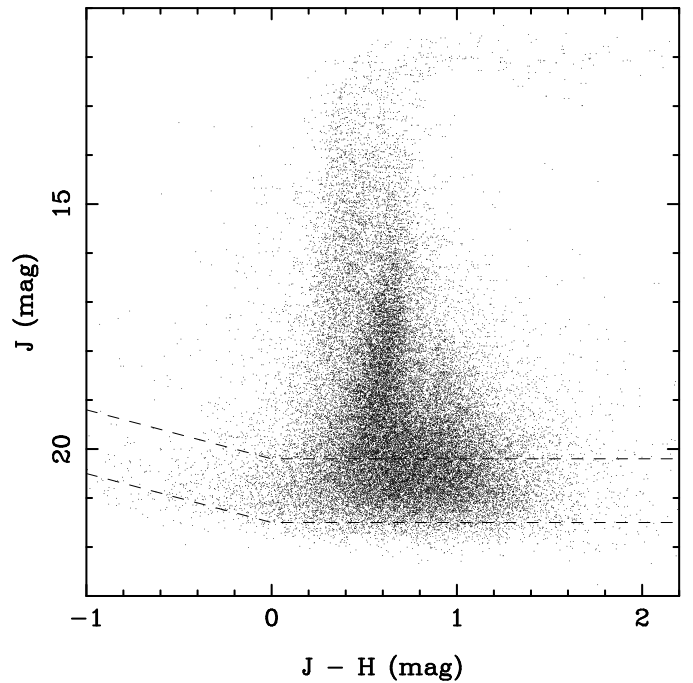}
   \caption{$J$ versus $J-H$ color-magnitude diagram resulting from the main survey that covers $\sim$0.8 deg$^2$ in the Pleiades cluster. Horizontal dashed lines stand for the search 10-$\sigma$ completeness and 3--4-$\sigma$ limiting magnitudes.}
              \label{alljh}
    \end{figure}

The $JH$ data of the main survey were cross-correlated using $\alpha$ and $\delta$ coordinates and a matching radius of 1\farcs5, resulting in over 43,000 sources in common to the first- and second-epoch observations.  All are plotted in the $J$ versus $J-H$ color-magnitude diagram illustrated in Figure~\ref{alljh}.

\subsection{Proper motions}
As indicated in Section~\ref{observations}, our images were calibrated in $\alpha$ and $\delta$ using UKIDSS GCS equatorial coordinates. In turn, UKIDSS GCS frames  were calibrated against bright stars of 2MASS; because these stars might have their own motions, it is possible that separate UKIDSS frames have varying astrometric zero-points. This uncertainty may amplify if UKIDSS data are used to calibrate other frames taken on different occasions. Consequently, we did not obtain proper motions via the direct subtraction of ($\alpha$, $\delta$) coordinates between epochs since this procedure may be subject to uncontrolled systematic effects. 

\begin{figure}
  \centering
  \includegraphics[width=0.48\textwidth]{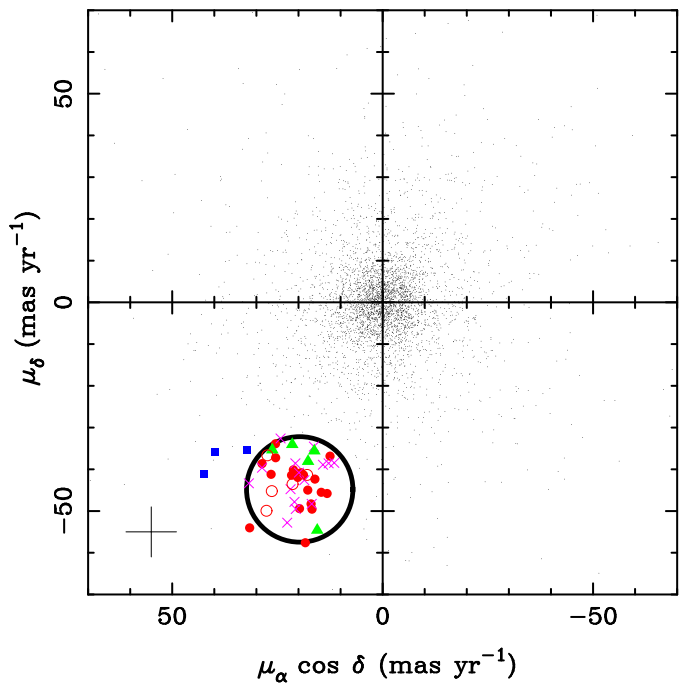}
  \caption{Proper motion diagram of the $JH$ survey. The mean motion of the Pleiades lies at the center  [(19.71, $-$44.82) mas\,yr$^{-1}$] of the black circle of radius of 13 mas\,yr$^{-1}$. The size of the circle corresponds to 1-$\sigma$ astrometric error bar of the faintest selected candidates. Red solid circles stand for unresolved cluster candidates with two or more proper
motion measurements; resolved sources are shown with green triangles; photometric cluster nonmembers are depicted with violet crosses; and the three faint objects complying with our photometric criteria found next to the astrometric circle of the Pleiades are shown with blue solid squares (see text). The tiny black dots around (0, 0) stand for a subsample of the entire volume of over 43,000 analyzed sources. The uncertainty associated with the proper
motion analysis is shown at the bottom-left side of the panel. \label{pm}}
  \end{figure}

For the fine proper motion analysis between the first-epoch data (older data, Omega-Prime) and the recent Omega-2000, HAWK-I, and LIRIS images, we compared pixel ($x$, $y$) coordinates frame per frame. We fit a third-order polynomial in $x$ and $y$ using IRAF tasks. A linear term and a distortion term were computed separately. The linear term included an $x$ and $y$ shift, an $x$ and $y$ scale factor,  a rotation, and a skew. The distortion term consisted of a polynomial fit to the residuals of the linear term. We found that the first-epoch $J$ data were rotated by about 0.7 deg with respect to the second-epoch $H$-band images. Between $\sim$100 and 150 sources per $J$-band frame were used to establish the polynomial fits; they yielded equation transformations with a mean uncertainty of $\pm$0\farcs054 ($\pm54$ mas, about a tenth of the Omega-2000 pixel) for the main survey. Considering the 9 yr baseline between first- and second-epoch data, this error translates into $\pm6$ mas\,yr$^{-1}$ in proper motion units. This mean that the uncertainty increases to $\pm$13 mas\,yr$^{-1}$ for sources with magnitudes close to the detection limit of the exploration. We determined this by quadratically combining the precision of the astrometric polynomial fits and the uncertainties associated with the calculation of the objects centroids. Figure~\ref{pm} shows the proper motion diagram resulting from the astrometric study of the main survey. The mean proper motion uncertainty associated with the comparison of the first $J$-band data and the very recent HAWK-I and LIRIS images is smaller because of the more optimal pixel scales of the latter instruments and the longer time intervals ($\sim$13 and $\sim$15 yr) between observations.

To improve the quality of our proper motion measurements, we also employed the recent $K_s$-band images acquired by the UKIDSS GCS project on 2009 August and 2010 October. UKIDSS $K_s$ images (5\farcm0\,$\times$\,5\farcm0 and 8\farcm0\,$\times$\,8\farcm0 in size) were downloaded from the archives with a pixel scale of 0\farcs2015, and were compared with the 1998 $J$-band Omega-Prime data, thus providing a time baseline of nearly 11 and 12 yr. Automatic object detection and proper motion analysis were conducted as previously explained, attaining astrometric precisions of $\pm$2.9 and $\pm$3.1 mas\,yr$^{-1}$ in $\mu_\alpha\,{\rm cos}\,\delta$ and $\mu_\delta$. UKIDSS photometry as provided by the catalog was also used to complement the near-infrared magnitudes of our data down to the $K_s$ detection limit of the GCS observations, which typically is $K_s \sim$ 18.4--18.7 mag at the 3 $\sigma$ level.

\section{Astrometric and photometric Pleiades candidates\label{astrometry}}
\begin{figure*}
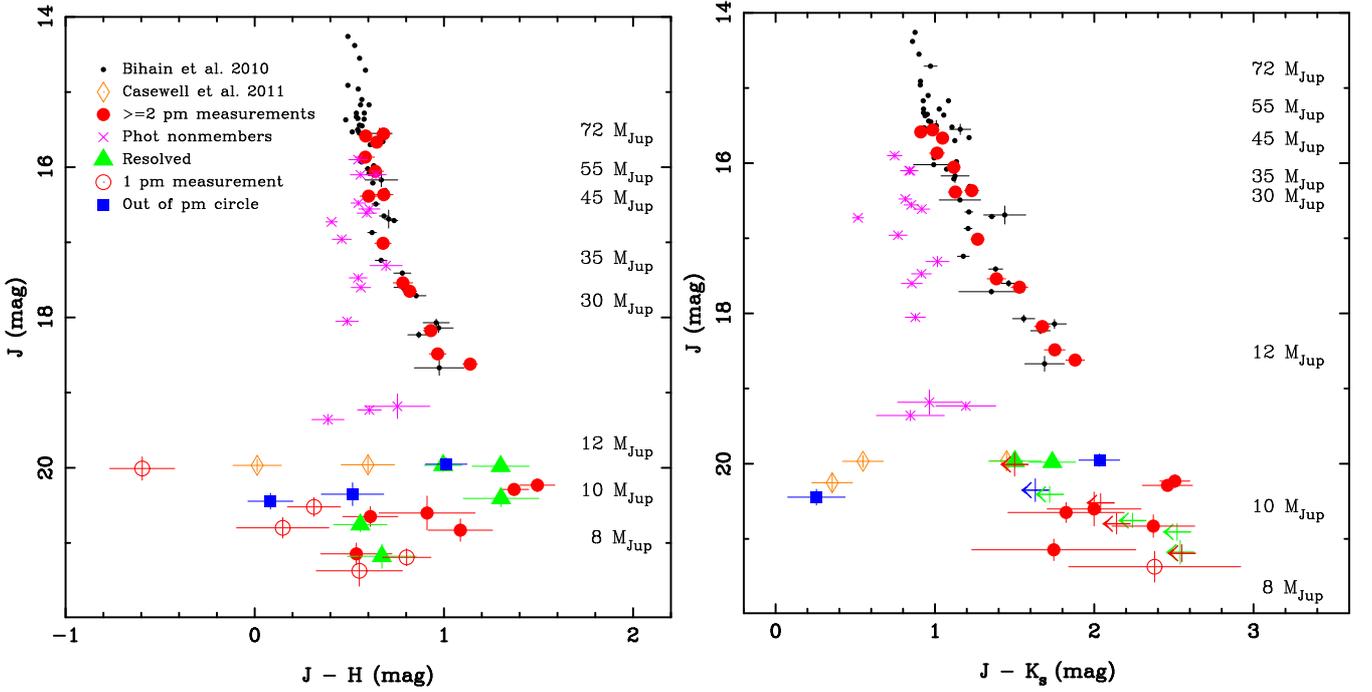

   \centering
   \includegraphics[width=0.48\textwidth]{jh.ps}     
   \includegraphics[width=0.48\textwidth]{jk.ps}
   \caption{Near-infrared color-magnitude diagrams of proper
motion selected objects. Nineteen red solid circles stand for unresolved Pleiades candidates with two or more proper motion measurements; five unresolved candidates with only one proper motion measurement (less reliable candidates) are depicted with red open circles. Five green triangles represent resolved sources; fifteen photometric cluster nonmembers are depicted with violet crosses; and the three faint objects complying with our photometric criteria found next to the astrometric circle of the Pleiades are shown with blue solid squares. The three unresolved T-type candidates of \citet{casewell11} are shown with brown diamonds.  Previously known proper motion Pleiades members that were confirmed spectroscopically are plotted as solid black dots (from the catalog by \citealt{bihain10}). In the right panel, UKIDSS $K_s$ nondetections are plotted as (color-coded) left arrows, indicating $J-K_s$ upper limits. Masses in Jupiter units are provided at the right side of the panels (based on the 120 Myr isochrone by \citealt{chabrier00a}).}
              \label{jhk}
    \end{figure*}

Cluster member candidates are selected in the magnitude range $J$\,=\,15.5--21.5 mag. The bright boundary corresponds to spectral type M6.5 and to the star--brown dwarf separation for the age, metallicity, and distance of the Pleiades \citep{stauffer98,martin98}. It also coincides with the so-called re-appearance of the lithium absorption line at 670.8 nm in the optical spectra of Pleiades brown dwarfs (\citealt{rebolo96}; see also the review by \citealt[and references therein]{basri00}). The faint boundary stands for the detection limit of the $JH$ main survey. Nevertheless, we claim entirety of identified Pleiades candidates from $J$ = 15.5 down to $\sim$20--20.3 mag, which is close to the completeness magnitude of the exploration. The final list of Pleiades candidates consists of sources that comply with the following astrometric and photometric criteria:

\begin{itemize}
\item In the proper motion diagram of Fig.~\ref{pm}, we identified all $JH$ sources with $\mu_\alpha$\,cos\,$\delta$ and $\mu_\delta$ motions contained within a circle of radius of 13 mas\,yr$^{-1}$ around the mean proper motion of the cluster [(19.71, $-$44.82) mas\,yr$^{-1}$, \citealt{lotkin03}]. The size of the searching circle matches the quoted proper motion error bar of the faint sources in the main survey.  Therefore, the astrometric search was performed at the 1-$\sigma$ level for $J \ge 20.3$ mag, and at the $\sim$2-$\sigma$ level for the bright magnitudes, indicating that Pleiades members at the faintest magnitudes of our survey might have been lost. On the one hand, opening the proper motion circle up to 2-$\sigma$ uncertainty at the faint end of the data would largely increase the number of non-Pleiades sources that
contaminate our survey. On the other hand, the searching proper
motion circle is greater in size than the velocity dispersion of Pleiades brown dwarfs reported in other studies ($\pm$7--10 mas\,yr$^{-1}$, \citealt{bihain06,casewell07}). We provide more details on the proper motion dispersion of Pleiades members in Section~\ref{pmdisp}.
\item The least massive Pleiades objects spectroscopically and astrometrically confirmed so far are classified in the near-infrared with types of  early-L through L4.5 \citep{bihain10}. They have $J \sim 17.5-18.8$ mag; consequently, the Pleiades photometric sequence is reasonably described down to these magnitudes \citep[and references therein]{bihain10,lodieu12}. Figure~\ref{jhk} shows the near-infrared color-magnitude diagrams of low-mass Pleiades members with spectra published in the literature (UKIDSS photometric system is used). Our newly selected proper motion candidates in the interval $J$ = 15.5--18.8 mag must follow the Pleiades sequence within the claimed photometric uncertainties and cluster  dispersion in all $JHK_s$ color-magnitude diagrams to be considered photometric cluster member candidates.
\item Beyond $J \sim 18.8$ mag and spectral type L4.5, the Pleiades sequence remains fully unknown. Discovering less luminous and cooler cluster sources is the main objective of this work. We did not assume any particular $J-H$ and/or $J-K_s$ color cuts; we allowed for both a natural extension of the known sequence toward redder indices and fainter magnitudes, and a possible sequence turn-over toward blue $J-H$ and/or $J-Ks$ colors, which would be expected if atmospheric methane absorption appears at near-infrared wavelengths. We left the proper motion study and the photometric criterion described next to reveal the true faint sequence of Pleiades members.
\item Because it is expected that true Pleiades members fainter than $J$ = 18.8 mag display spectral types cooler than L4.5, we imposed red optical-to-near-infrared color cuts based on field dwarfs with UKIDSS photometry (see \citealt{hewett06}) and the previous works on the Pleiades by \citet{lodieu07,lodieu12}, i.e., $Z-J \ge 2.3$ mag, or nondetection in the UKIDSS $Z$-band. 
\item Finally, all candidates were visually inspected, and all images available to us were blinked to check that there is a motion in the direction of the Pleiades to the naked eye.
\end{itemize}


   \begin{figure}
   \centering
   \includegraphics[width=0.48\textwidth]{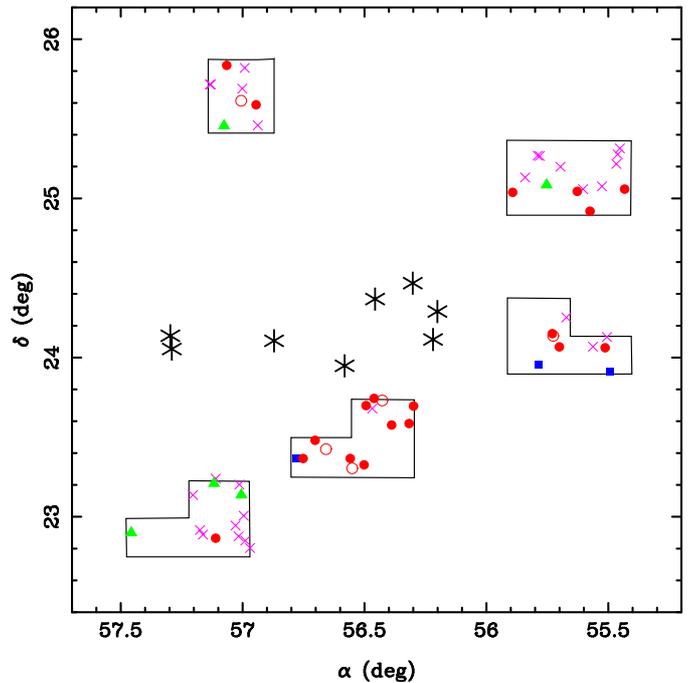}     
   \caption{Location of the selected objects within the surveyed area of $\sim$0.8 deg$^2$ (black rectangles). Asterisks stand for eight well-known Pleiades stars. Other symbols are as in Figs.~\ref{pm} and \ref{jhk}.}
              \label{cand_loc}
    \end{figure}
%


%


\addtocounter{table}{1}
\begin{table*}
\caption{Other proper-motion-selected objects. \label{cands2}}
\scriptsize
\begin{tabular}{lccccccc}
\hline\hline
\multicolumn{7}{l}{Unresolved, photometric nonmembers ($J=15.5-19.5$ mag).} \\
\hline
Object &  RA, DEC (J2000) & $\mu_\alpha$ cos $\delta$, $\mu_\delta$ & $N$\tablefootmark{a} & $J$\tablefootmark{b} & $H$\tablefootmark{b} & $K_s$\tablefootmark{c} & $\sigma (\mu_\alpha, \mu_\delta)$\tablefootmark{d} \\
       & ($^{\rm h}$ $^{\rm m}$ $^{\rm s}$),  ($^{\circ}$ $'$ $''$) & (mas\,yr$^{-1}$) & & (mag) & (mag) & (mag) & (mas\,yr$^{-1}$)   \\
\hline
PNM01  &  3 48 41.99  $+$22 54 58.2 &  18.6 $\pm$ 5.7,   $-$42.6 $\pm$ 6.6 &   2 &  15.90 $\pm$ 0.04 & 15.35 $\pm$ 0.02 &   15.15 $\pm$ 0.01 &             \\ 
PNM02  &  3 41 50.77  $+$25 16 34.9 &  21.0 $\pm$ 5.7,   $-$47.8 $\pm$ 6.6 &   1 &  16.10 $\pm$ 0.05 & 15.56 $\pm$ 0.01 &   15.26 $\pm$ 0.01 &             \\ 
PNM03  &  3 48 07.02  $+$22 56 39.7 &  28.7 $\pm$ 5.7,   $-$39.6 $\pm$ 6.6 &   2 &  16.10 $\pm$ 0.04 & 15.54 $\pm$ 0.03 &   15.26 $\pm$ 0.01 &             \\ 
PNM04  &  3 48 31.36  $+$25 42 57.1 &  20.7 $\pm$ 5.7,   $-$38.6 $\pm$ 6.6 &   2 &  16.48 $\pm$ 0.03 & 15.93 $\pm$ 0.01 &   15.66 $\pm$ 0.02 &             \\ 
PNM05  &  3 48 03.19  $+$23 12 05.8 &  22.7 $\pm$ 5.7,   $-$52.8 $\pm$ 6.6 &   2 &  16.56 $\pm$ 0.04 & 15.95 $\pm$ 0.03 &   15.71 $\pm$ 0.02 &             \\ 
PNM06  &  3 42 06.21  $+$25 04 31.0 &  16.3 $\pm$ 5.8,   $-$34.6 $\pm$ 6.8 &   2 &  16.61 $\pm$ 0.04 & 16.02 $\pm$ 0.01 &   15.70 $\pm$ 0.01 &             \\ 
PNM07  &  3 41 51.98  $+$25 13 00.5 &  12.9 $\pm$ 5.8,   $-$38.6 $\pm$ 6.8 &   2 &  16.73 $\pm$ 0.02 & 16.32 $\pm$ 0.01 &   16.21 $\pm$ 0.02 &             \\ 
PNM08  &  3 48 32.26  $+$25 42 57.0 &  19.9 $\pm$ 5.8,   $-$40.8 $\pm$ 6.8 &   2 &  16.96 $\pm$ 0.04 & 16.50 $\pm$ 0.01 &   16.19 $\pm$ 0.03 &             \\ 
PNM09  &  3 43 07.44  $+$25 15 59.9 &  31.8 $\pm$ 5.9,   $-$43.4 $\pm$11.0 &   2 &  17.31 $\pm$ 0.07 & 16.62 $\pm$ 0.04 &   16.29 $\pm$ 0.02 &             \\ 
PNM10  &  3 47 58.00  $+$25 49 12.5 &  24.3 $\pm$ 5.9,   $-$32.6 $\pm$ 7.0 &   2 &  17.47 $\pm$ 0.04 & 16.93 $\pm$ 0.01 &   16.56 $\pm$ 0.04 &             \\ 
PNM11  &  3 48 03.95  $+$22 52 37.2 &  16.9 $\pm$ 5.9,   $-$48.3 $\pm$ 7.0 &   3 &  17.60 $\pm$ 0.03 & 17.04 $\pm$ 0.03 &   16.75 $\pm$ 0.05 &  1.0, 2.5   \\ 
PNM12  &  3 42 47.10  $+$25 11 57.1 &  11.5 $\pm$ 6.0,   $-$38.5 $\pm$ 7.1 &   2 &  18.05 $\pm$ 0.04 & 17.56 $\pm$ 0.03 &   17.18 $\pm$ 0.04 &             \\ 
PNM13\tablefootmark{e}  &  3 43 09.51  $+$25 16 06.1 &  20.7 $\pm$ 8.0,   $-$49.5 $\pm$ 9.0 &   2 &  19.18 $\pm$ 0.16 & 18.43 $\pm$ 0.05 &   18.22 $\pm$ 0.11 &             \\ 
PNM14  &  3 48 00.32  $+$25 41 24.3 &  21.9 $\pm$ 8.0,   $-$44.8 $\pm$ 9.0 &   3 &  19.23 $\pm$ 0.05 & 18.63 $\pm$ 0.03 &   18.04 $\pm$ 0.18 &  9.9, 2.0   \\ 
PNM15  &  3 48 39.08  $+$22 53 15.9 &  14.2 $\pm$ 8.0,   $-$38.9 $\pm$ 9.0 &   3 &  19.36 $\pm$ 0.07 & 18.97 $\pm$ 0.05 &   18.51 $\pm$ 0.20 & 20.0, 4.4   \\ 
\hline
\hline
\end{tabular}
\begin{tabular}{lccccccc}
\multicolumn{8}{l}{Unresolved objects with $J \ge 20$ mag and a proper motion larger than the explored astrometric circle.} \\
\hline
Object &  RA, DEC (J2000) & $\mu_\alpha$ cos $\delta$, $\mu_\delta$ & $J$\tablefootmark{b} & $H$\tablefootmark{b} & $K_s$\tablefootmark{c} & $W1$ & $W2$ \\
       & ($^{\rm h}$ $^{\rm m}$ $^{\rm s}$),  ($^{\circ}$ $'$ $''$) & (mas\,yr$^{-1}$) & (mag) & (mag) & (mag) & (mag) & (mag)  \\
\hline
NPM01\tablefootmark{e} & 3 41 58.20  $+$23 54 44.1 &    39.8 $\pm$ 10.8,  $-$35.9 $\pm$ 10.1 &   19.95 $\pm$ 0.08 & 18.94 $\pm$ 0.03 &  17.92 $\pm$ 0.10 &  16.80 $\pm$ 0.10 & 16.46 $\pm$ 0.27 \\ 
NPM02 & 3 43 08.58  $+$23 57 10.0 &    32.2 $\pm$  9.6,  $-$35.4 $\pm$  9.8 &   20.35 $\pm$ 0.15 & 19.84 $\pm$ 0.06 &  $>$18.72         &                   &                  \\ 
NPM03 & 3 47 07.35  $+$23 21 56.0 &    42.4 $\pm$ 11.7,  $-$41.0 $\pm$ 10.7 &   20.44 $\pm$ 0.10 & 20.36 $\pm$ 0.06 &  20.19 $\pm$ 0.15\tablefootmark{f} &                   &                  \\ 
\hline
\end{tabular}
\\
\tablefoottext{a}{Number of proper motion measurements.}
\tablefoottext{b}{This paper.}
\tablefoottext{c}{From UKIDSS GCS, with a few indicated exceptions.}
\tablefoottext{d}{Standard deviation of the mean proper motion measurement computed for $N \ge 3$.} 
\tablefoottext{e}{Close to $J$- and/or $H$-band frame border, large uncertainty in both photometric and astrometric data.}
\tablefoottext{f}{LIRIS $K_s$ measurement.}
\end{table*}

A total of 44 sources of the $JH$ main survey pass the astrometric criterion. They are shown separated into different groups, as explained next, in the  ($\alpha$, $\delta$) plane (Figure~\ref{cand_loc}) and in the $J$ versus $J-H$, $J-K_s$ color-magnitude diagrams (Figure~\ref{jhk}). Out of 44, 29 (Table~\ref{cands}) comply with all the photometric requirements. Nineteen have two or more proper motion measurements or have been previously identified in other studies \citep{pinfield00,nagashima03,lodieu12}, that
is to say, they have been detected and selected on various occasions. This group of 19 represents the list of unresolved candidates with the highest probability of being genuine Pleiades members. In the top panel of Table~\ref{cands} we provide the derived $\alpha$ and $\delta$ coordinates from the 2007 $H$-band images, proper motions (equally weighted mean values whenever two or more measurements are available), the number of proper motion measurements ($JH$ main survey, data coming from UKIDSS GCS $K_s$ images, and follow-up data obtained by us as indicated in Section~\ref{observations}), and the standard deviation of the proper motions if there are three or more astrometric measures. The low values of the proper
motion standard deviations suggest that the quoted astrometric error bars are likely overestimated particularly for the bright magnitudes. Table~\ref{cands} also lists the names for all newly proposed substellar candidates after \citet{stauffer94}, \citet{rebolo95}, and \citet{osorio97}, meaning that names agree with the observatory in which they were observed, followed by the word Pleiades, and numbered according to their increasing $J$-band apparent magnitude. The last column of the top panel of Table~\ref{cands} indicates other names for known objects. The UKIDSS GCS survey of \citet{lodieu12} has covered the entire area of our exploration, yet about 31\%~of the 13 candidates between $J$ = 15.5 and 18.8 mag were missed in their work, probably because of the different approach in the analysis of proper motions and the distinct color-magnitude diagrams for object selection employed by the two groups.

Of the 29 photometric and astrometric candidates, five ($J \ge 20$ mag) appear to be unresolved and have one proper motion measurement because they are not detected in any passband of UKIDSS GCS and are not contained in any of our follow-up images. Another five (also among the faintest candidates) appear to have slightly resolved FWHMs in the $H$-band frames, which are the deepest images in all explored data. These two groups are listed in the middle and bottom panels of Table~\ref{cands}. We remark that for $J \ge 20.5$ mag, even point-like candidates may be extragalactic objects that are not spatially resolved at the seeing of our observations or are too faint and have uncertain FWHMs particularly in the $J$- and $K_s$-bands.

Of the original 44 astrometric candidates, 15 are photometric nonmembers since their $J-H$ and/or $J-K_s$ colors deviate from the known Pleiades sequence, or they have blue UKIDSS $Z-J$ colors. Nevertheless, we list them in the top panel of Table~\ref{cands2} for future works.

Complementary to the explored astrometric circle depicted in the ($\mu_\alpha$\,cos\,$\delta$, $\mu_\delta$) plane of Fig.~\ref{pm}, we also investigated faint sources with $J \ge 20$ mag, $\mu_\delta$ motions similar to that of the Pleiades cluster, and $\mu_\alpha$\,cos\,$\delta$ displacements beyond the circle toward the east and up to 2-$\sigma$ the corresponding astrometric uncertainty. Three objects were found and are listed in the bottom panel of Table~\ref{cands2}. These data were used to estimate possible contaminants at the faintest magnitudes as well as the feasibility of the survey to detect very dim sources with near-infrared colors from blue to very red indices.

For all 44 sources, our $JH$ and cataloged UKIDSS $K_s$ photometry (when available) is given in Tables~\ref{cands} and~\ref{cands2}. In case of $K_s$ nondetections, we provide the limiting magnitudes at the 3 $\sigma$ level computed for each individual UKIDSS GCS frame. For two objects, Calar Pleiades 21 (hereafter Calar~21) and NPM03, we derived the $K_s$ magnitudes using LIRIS data obtained on 2013 October 11. Calar~26 (the faintest object in the top panel of Table~\ref{cands}) was not detected automatically by the UKIDSS GCS pipeline; however, visual inspection of the downloaded $K_s$ image reveals a faint source near the location of our candidate. The measured Omega-prime versus UKDISS GCS proper motion of Calar~26 agrees with the main survey value within the astrometric uncertainties. LIRIS and UKIDSS $K_s$ photometry of Calar~21, 26, and NPM03 was derived as described in Section~\ref{phot}, and is provided in Tables~\ref{cands} and~\ref{cands2}.

Additionally, the $\alpha$ and $\delta$ coordinates of our objects were cross-correlated against the {\sl WISE} catalog \citep{wright10} to obtain their mid-infrared colors. {\sl WISE} data are given in Tables~\ref{cands} and~\ref{cands2}. We used a cross-correlation radius of 2\arcsec~and considered $W1$ (3.3526 $\mu$m) and $W2$ (4.6028 $\mu$m) magnitudes for which the catalog claims detections with signal-to-noise (S/N) ratios of 3 and higher. The great majority of our sources are not detected at longer wavelengths. Note that despite the generous matching radius, the differences in the equatorial coordinates between the $JH$ and {\sl WISE} sources are always smaller than 1\arcsec, except for Calar~26, where we found a difference of 1\farcs99, thus suggesting that the {\sl WISE} source may not be the same as the $JH$ one. We exclude Calar~26 from the color-magnitude diagrams involving {\sl WISE} data throughout this work.

\section{Pleiades photometric sequence\label{seq}}

\subsection{Contaminants}
Before proceeding with the description of the Pleiades photometric sequence properties and mass function, we need to discuss the expected object contamination.   Our own survey can be used as the basis for studying the number of possible contaminants throughout all the magnitude intervals of interest.  In Fig.~\ref{jhk}, between the substellar limit ($J = 15.5$ mag) and the completeness magnitude of the main survey ($J \sim 20.2$ mag), there are 13 objects with proper motion and photometry consistent with Pleiades membership, and 15 sources whose photometry is not compatible with that of the cluster. This suggests that the initial astrometric sample of unresolved objects suffers from a contamination of about 50\%~in the magnitude interval $J \approx 15.5-20$ mag. Note that this contamination fraction results from the multiwavelength inspection of the data: Within this magnitude range, bona fide substellar Pleiades members show well-constrained red $J-H$ and $J-K_s$ colors  typical of mid-M through late-L types, while the astrometric contaminants display bluer colors typical of early- and mid-M types. 
 
By extrapolating the astrometric contamination fraction of $\sim$50\%~below the survey completeness, of the initial 11 astrometric, unresolved candidates with $J \approx 20-21.5$ mag only about 5--6 would be true Pleiades members and the other half would be unrelated to the cluster. This contaminating count coincides with the number of faint, near-infrared blue and red sources populating regions immediately adjacent to the astrometric circle described in Sect.~\ref{astrometry} and shown in Fig.~\ref{pm} (the three actual detections of the region to the east of the astrometric circle must be conveniently scaled by a factor of 1.5 to account for the ratio between the two explored proper motion areas). Furthermore, a similar count ($\sim$3) of L5--T5 field dwarf contaminants would be expected in the range of $J =20-21.5$ mag and in a survey like ours that
covers an area of 1 deg$^2$ according to the recipes by \citet{caballero08}. Note that all of our faint candidates show colors typical of L and T dwarfs. Therefore, we used 13 and 6 tentative cluster members in the intervals $J \approx 15.5-20$ and $\approx 20-21.5$ mag, respectively, to derive the cluster mass function in Sect.~\ref{massf}.

All 11 astrometric point-like candidates with $J \approx 20-21.5$ span a wide range of $J-H$ colors (Figure~\ref{jhk}). Seven of them have $K_s$-band detections displaying quite red $J-K_s$ colors, and for six out of seven (see Table~\ref{cands}) we were able to confirm their $JH$ proper motion measurements via third-epoch images. We have least confidence that the five unresolved candidates with just one proper motion derivation are reliable Pleiades member candidates. More follow-up data are required to confirm their cluster membership status. In contrast, the six unresolved sources with $K_s$ detections in UKIDSS and LIRIS are  noteworthy because their proper motions are double-checked, which makes these objects very promising low-mass Pleiades candidates. Furthermore, they have photometric properties that differ from those of the field in a way described in the following sections. We rely on these six sources to characterize the Pleiades photometric sequence at very faint magnitudes.

Extragalactic objects may also be a source of contamination in our survey, particularly at the faint end, where the proper motion uncertainties are large and the extragalactic population increases significantly. Given the good seeing of some data, we were able to identify five (out of a total of 16) possibly resolved objects in the interval $J \approx 20-21.5$, which are not considered further in this paper. As discussed in \citet{bihain09}, mid-L to mid-T type dwarfs are distinguishable from the great majority of galaxies in optical-to-near-infrared color-color diagrams since the dwarfs have very red colors, while the galaxies tend to be bluer. By placing the six $J \approx 20-21.5$ mag UKDISS and LIRIS unresolved candidates in the color-color and color-magnitude diagrams of \citet{bihain09}, which include the galaxies from the GOODS-MUSIC catalog \citep{grazian06}, we observed that most seem to fall in regions of very low density extragalactic population. Furthermore, our candidates with the reddest $J-K_s$ colors are very likely Galactic sources since they tend to fall at indices redder than those of the GOODS-MUSIC galaxies. Only follow-up spectroscopy will be able to discern how many galaxies there are in our final list of unresolved faint Pleiades candidates of Table~\ref{cands}.

   \begin{figure*}
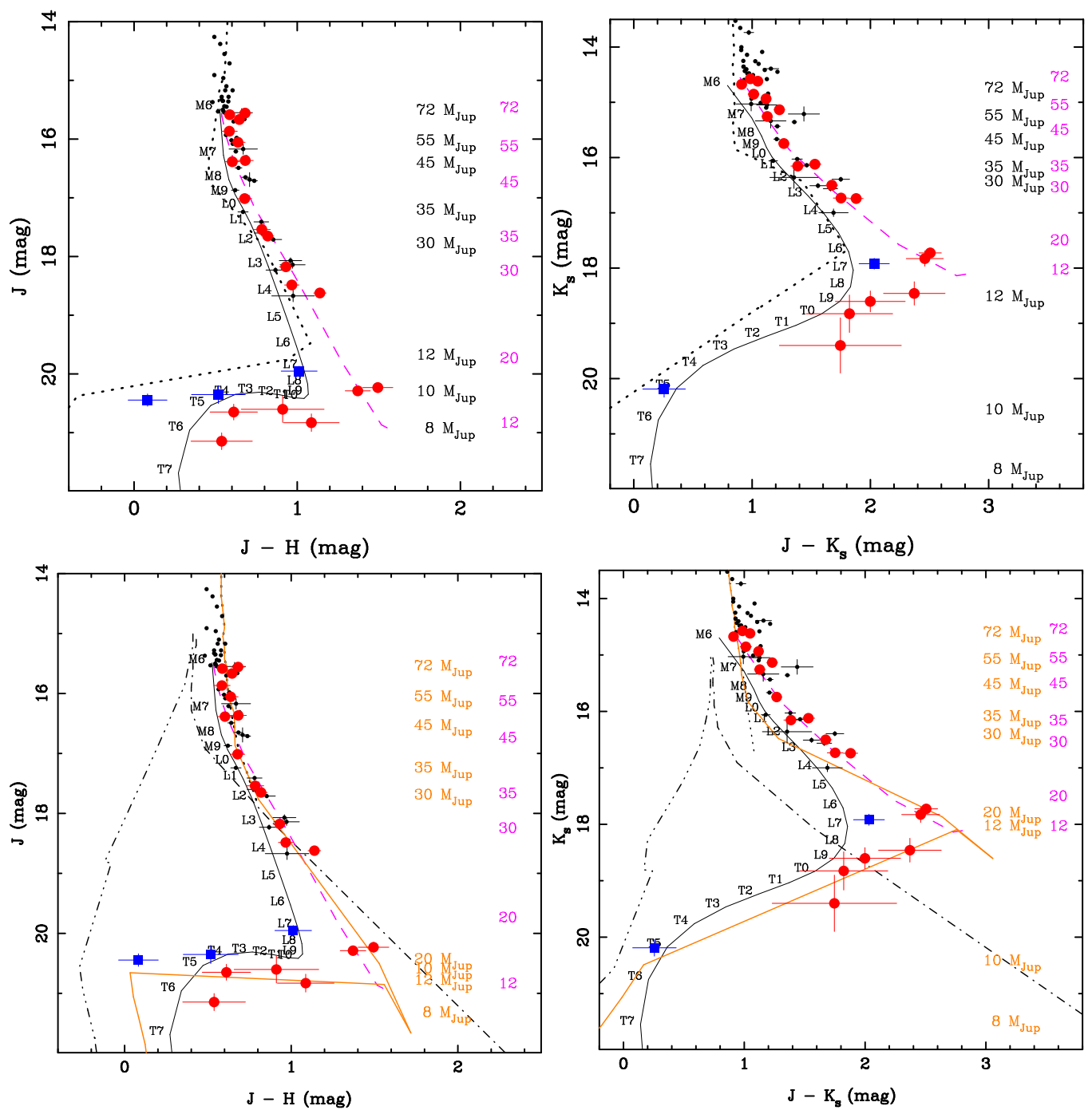

   \centering
   \includegraphics[width=0.48\textwidth]{jh2.ps}     
   \includegraphics[width=0.48\textwidth]{jk3.ps}
   \includegraphics[width=0.48\textwidth]{jh2nextgen.ps}
   \includegraphics[width=0.48\textwidth]{jk3nextgen.ps}
   \caption{Near-infrared color-magnitude diagrams of selected Pleiades candidates with two or more proper motion measurements. Symbols are as in Fig.~\ref{jhk}. The solid line corresponds to the mean observed sequence of standard field late-M, L, and T dwarfs \citep{dupuy12} taken to the distance of the Pleiades. Spectral types are labeled. We show a semi-empirical, solar-metallicity, 120 Myr isochrone obtained as explained in the text (based on models by \citealt{chabrier00a}) with a magenta dashed line; related masses are also labeled in magenta in all panels. In the top panels, the BT-Settl 120 Myr isochrone based on synthetic colors \citep{chabrier00a,allard12} is shown with a dotted line. Masses based on this model are provided on the right side in black. In the bottom panels, the NextGen (dotted line), DUSTY (dot-dashed line), and COND (dot-dot-dashed line) observable magnitudes are illustrated. The combined NextGen$+$DUSTY$+$COND 120 Myr evolutionary track (see text) is depicted with a brown solid line; masses given by this model are also indicated in brown. } 
              \label{jhk2}
    \end{figure*}

\subsection{Color-magnitude diagrams}
Various combinations of near- and mid-infrared color-magnitude diagrams are depicted in Figs.~\ref{jhk2} and~\ref{wise}, where only the 13$+$6 most promising Pleiades member candidates are shown; these are the unresolved candidates with two or more proper motion measurements. The sequences defined by these objects become increasingly red from $J = 15.5$ mag down to $J \sim 20.3$ mag, thus indicating that any color turn-over of the Pleiades sequence must occur beyond $J \sim 20.3$. The reddest $J-H$, $J-K_s$, $J-W1$, and $J-W2$ indices of cluster member candidates are measured at $\sim$1.5, $\sim$2.5, $\sim$3.8, and $\sim$5.0 mag, respectively. With the exception of the $K_s$ versus $K_s-W2$ diagram of Fig.~\ref{wise}, fainter cluster candidates appear to have bluer indices, particularly those involving the $J$ band, suggesting that either they are contaminants or the color-magnitude disposition of Pleiades members turns to blue colors at $J \ge 20.3$ mag. To confirm this finding, follow-up spectroscopic observations need to be obtained to determine the nature of our candidates. 

The very red colors of our $J \sim 20.2$-mag candidates contrast with the objects discovered by \citet{casewell07,casewell10}, who found tentative proper motion T-type Pleiades members with blue $J-H$ colors at about $J \sim 20.2-20.3$ (see Figure~\ref{jhk}). A few of their most interesting "blue" candidates were followed-up photometric and spectroscopically with methane filters from the ground, and with {\sl Spitzer} and the Hubble Space Telescope (HST) from space by \citet{casewell11} and \citet{lucas13}. None appears to have methane absorption, their mid-infrared colors are unexpected for normal dwarfs, and the HST image of the Pleiades candidate PLZJ\,93 shows a diffuse halo, demonstrating that it is not a T dwarf;  it very likely is an external galaxy. We remark that extragalactic objects constitute one critical source of contamination in very deep surveys like the one carried out by \citet{casewell07} and ours.

Another striking property of the photometric sequences shown in Figures~\ref{jhk2} and~{\ref{wise} is the apparent lack of candidates or a less populated magnitude domain in the interval $J \sim 18.8-20.2$ mag, in contrast to the nice and homogeneous object continuity above and below this signature. Furthermore, this apparent hole falls within the completeness magnitude of our survey; we thus conclude that it is a real feature of the Pleiades substellar sequence. This attribute significantly reduces in size or becomes even imperceptible at longer wavelengths (e.g., $K_s$ and the {\sl WISE} filters), probably because of the dramatic change toward redder colors of cluster member candidates with $J > 18.5$ mag. As is illustrated in the $K$-band panels of Figures~\ref{jhk2} and~\ref{wise}, this feature in $K_s$ appears to lie in the interval 17--17.7 mag, which is half as large as that of the $J$ band, and no obvious discontinuity is observed in the $W1$ or $W2$ magnitudes.

The physical origin of this property is yet to be explained; however, it may be related to the increasing presence of dust in the upper photospheric layers of ultracool dwarfs with low-gravity (or low-pressure) atmospheres and decreasing effective temperature, and/or the possible existence of warm dusty (debris) disks (or rings) very near the central objects, which would make Pleiades low-mass substellar members appear very red and somehow ``extinguished'' at 1 $\mu$m (see next). The current knowledge of the formation and evolution of debris disks around substellar sources and their relation to the central objects properties is very scarce; therefore, this possibility, although less explored in the literature than the atmospheric dust scenario, cannot be discarded. These two views have been discussed in other studies to account for the very red near-infrared colors of "young" field dwarfs (e.g., \citealt[and references therein]{osorio10,gizis12,faherty12,faherty13}).

\subsection{Comparison with the field}

   \begin{figure*}
   \centering
   \includegraphics[width=0.48\textwidth]{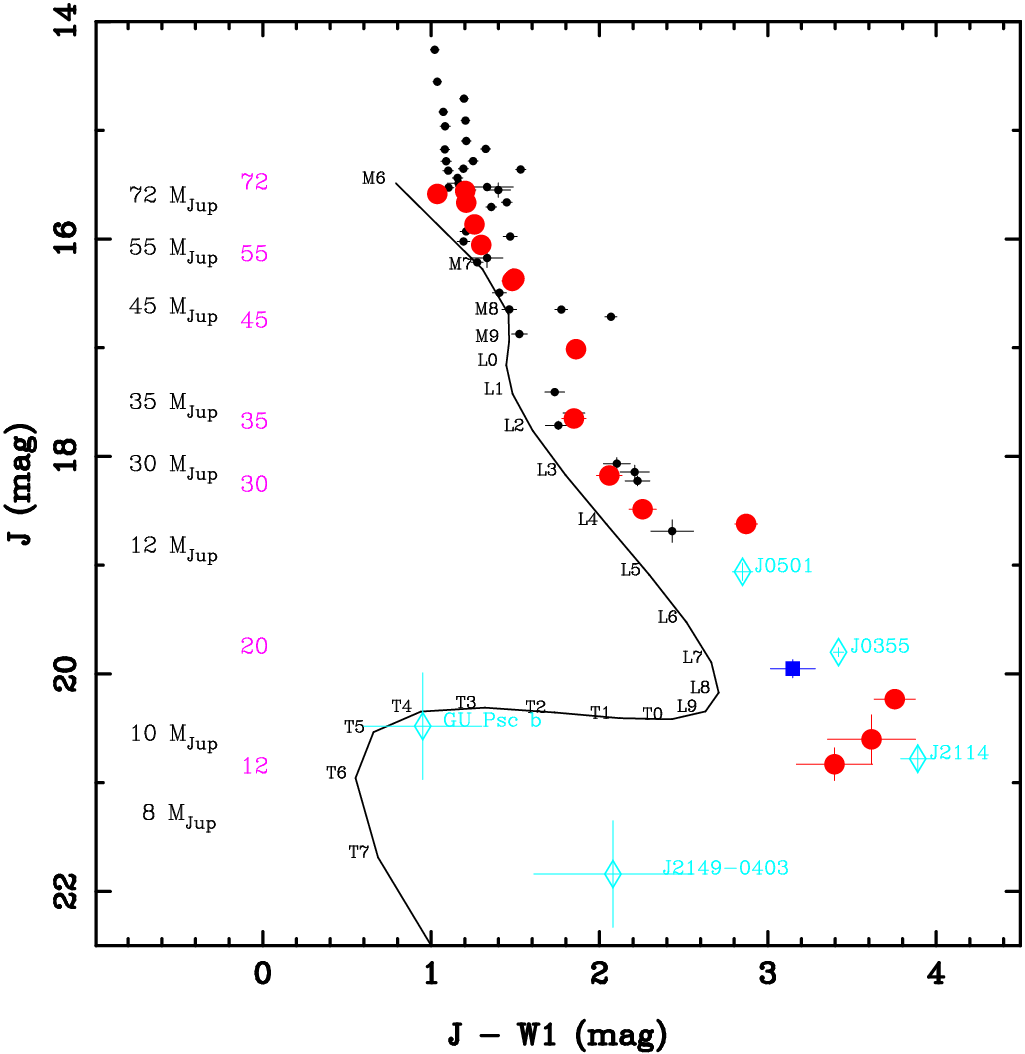}
   \includegraphics[width=0.48\textwidth]{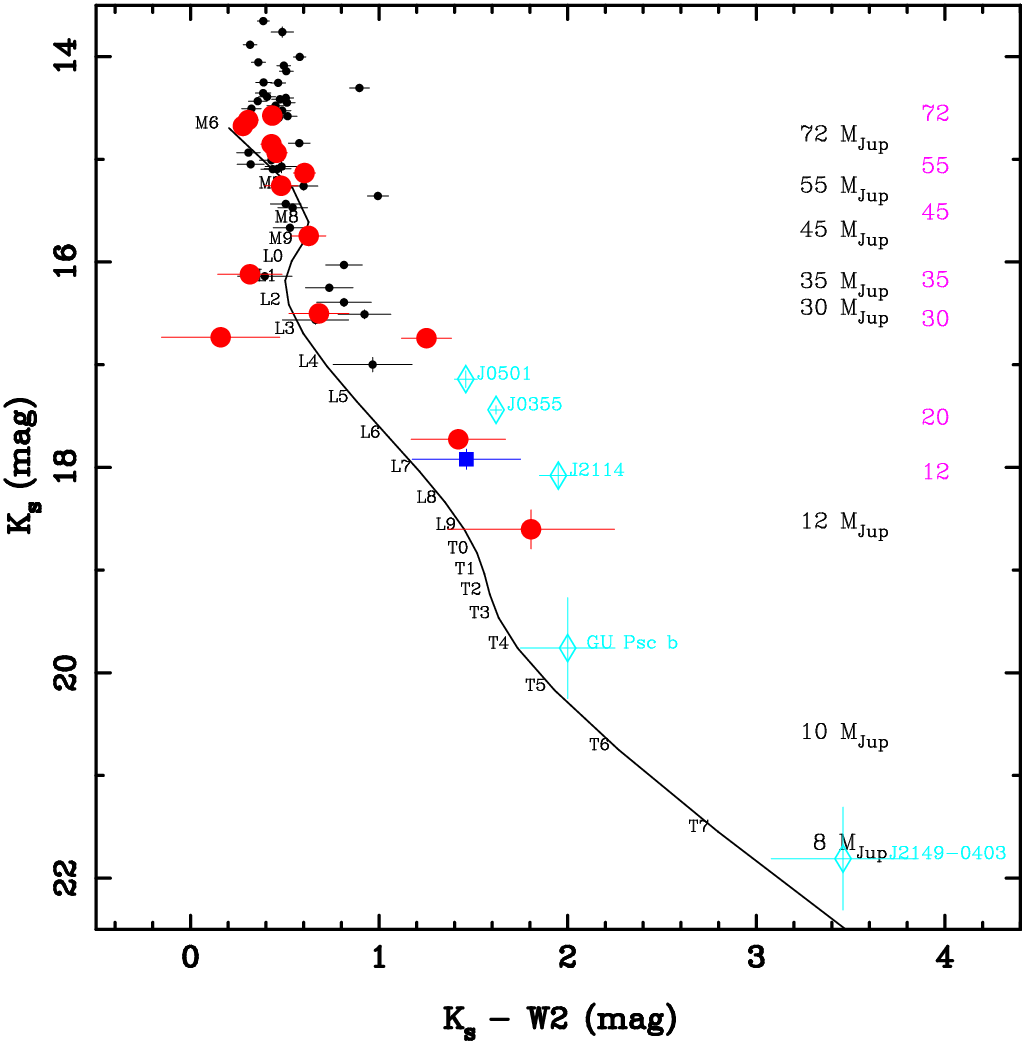}     
   \includegraphics[width=0.48\textwidth]{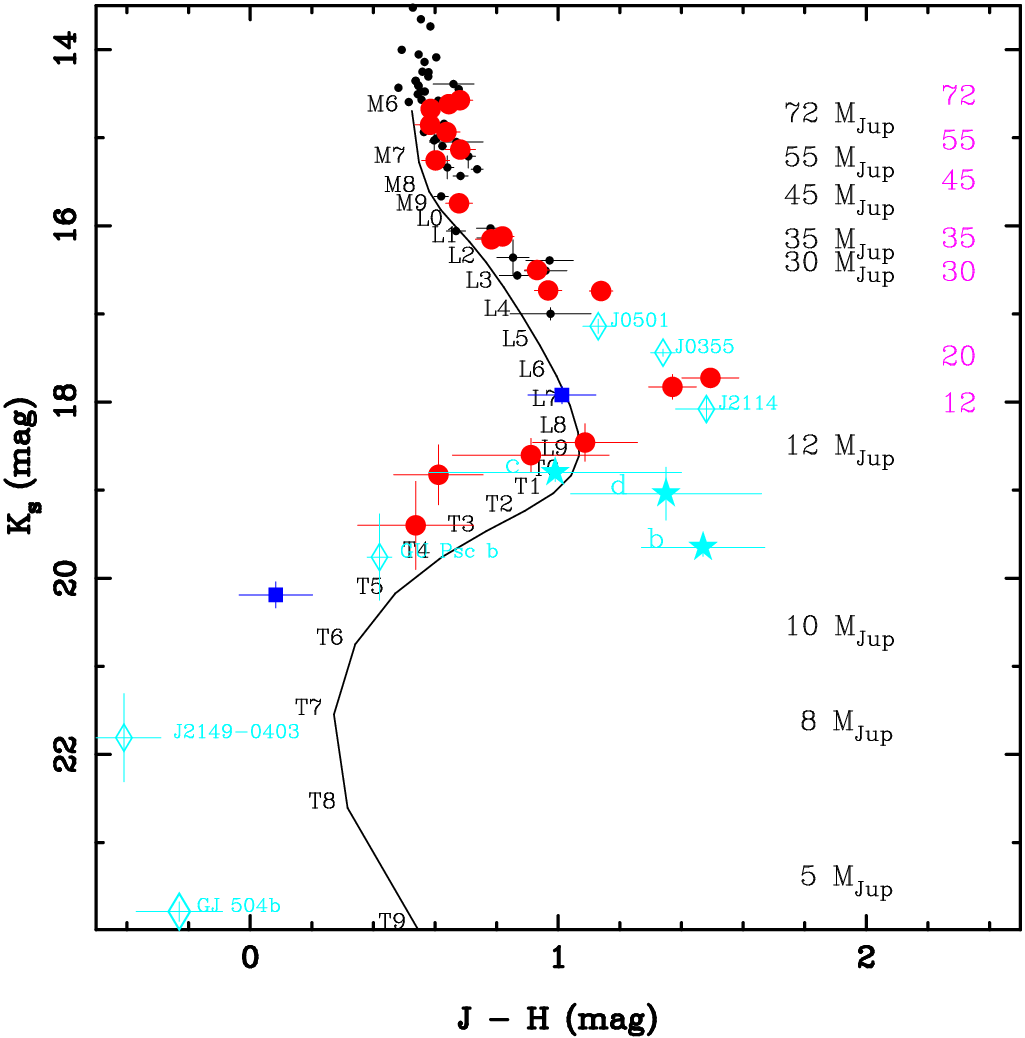}
   \includegraphics[width=0.48\textwidth]{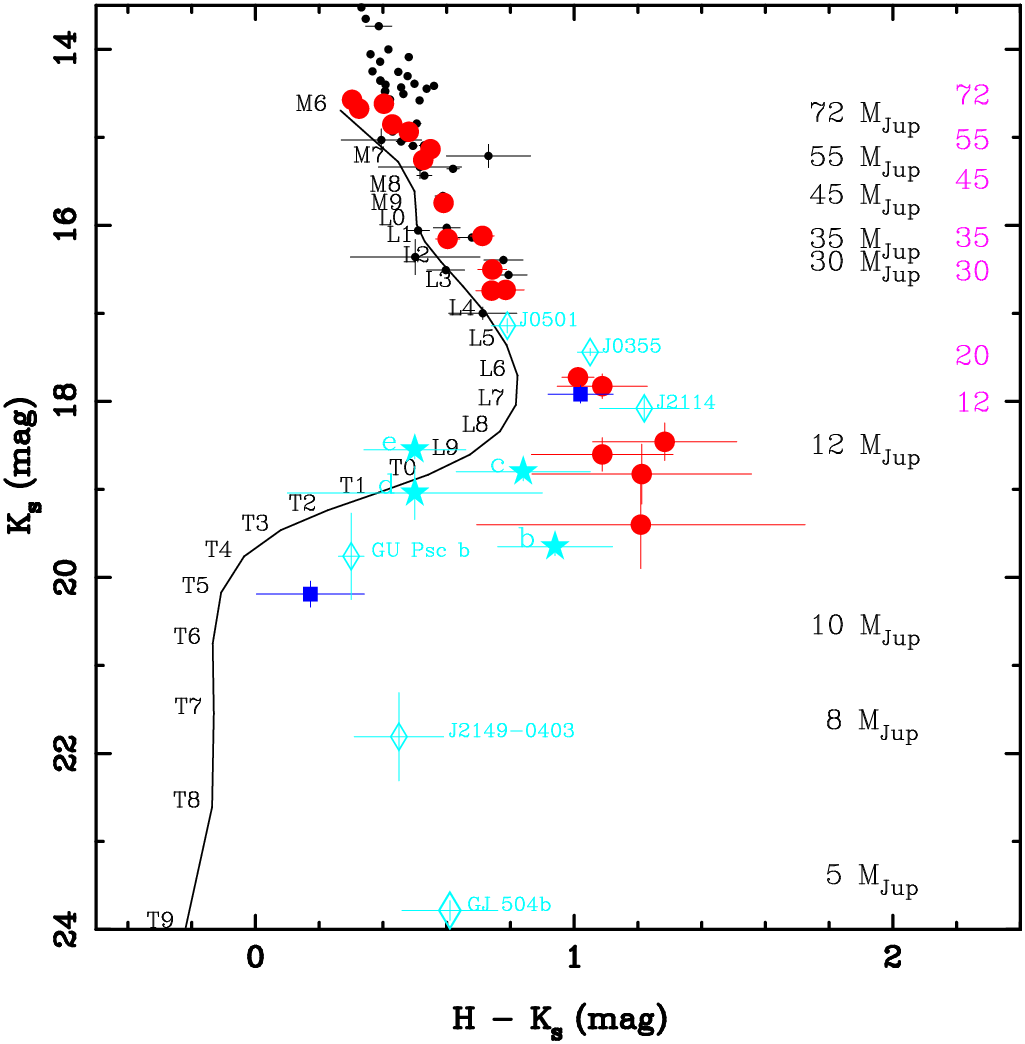}     
   \caption{Near-infrared and {\sl WISE} photometry of selected Pleiades candidates with two or more proper motion measurements. Symbols are as in Fig.~\ref{jhk}. The field "young" L and T dwarfs (\citealt{cruz09,delorme12,liu13,kuzuhara13,naud14}) are denoted with open diamonds. The star symbols stand for the bcde planets orbiting HR\,8799 \citep{marois08,marois10}. The solid line corresponds to the mean observed sequence of field late-M, L, and T dwarfs \citep{dupuy12} moved to 133.5 pc. All field objects are also taken to the Pleiades distance. Spectral types and objects from the literature are labeled. The {\sl WISE} photometry of J2149$-$0403 is quite uncertain; \citet{delorme12} argued that the $W2$ detection lied at the exact expected position, while the $W1$ detection was slightly offset. Masses are given in Jovian units color-coded as in Fig.~\ref{jhk2}. }
              \label{wise}
    \end{figure*}

The defined Pleiades photometric sequence is compared with the sequence of field M, L, and T-type dwarfs in Fig.s~\ref{jhk2} and~\ref{wise}. For the field we adopted the polynomial fits of Mauna Kea Observatory and {\sl WISE} absolute magnitudes as a function of spectral type for all bands given in Table~14 by \citet{dupuy12}. These fits have a typical scatter of $\pm$0.4 mag. Absolute magnitudes were taken to the distance of the cluster. Based on the colors of field dwarfs, our objects have near- and mid-infrared indices that cover spectral types from $\sim$M6.5 through early-Ts. Globally, the field sequence seems to follow the general trend described by the Pleiades candidates in all color-magnitude diagrams.

However, we observe some discrepancies. From $J, K_s = 15.5, 14.5$ down to 18.8, 16.8 mag, Pleiades members appear slightly overluminous in $J$ and $K_s$ with respect to the field, as expected for their young age (according to the theory of substellar evolution, Pleiades low-mass objects must be undergoing self-gravitational contraction). \citet{bihain10} discussed that these objects, which have late-M and mid-L types, appear modestly redder in the near-infrared than  their spectral counterparts in the field. Cluster member candidates with $J \sim 20.2-20.3$ and $K_s \sim  17.8$ mag (Calar~21 and~22) also show up redder than the field in all $J-H$, $H-K$, $J-K$, and $J-W1$ colors by significant amounts of 0.5--1 mag. Fainter candidates display an heterogeneous behavior in the various near- and mid-infrared color-magnitude diagrams: While they emerge redder than the field in $H-K$, $J-K$, and $J-W1$, their $J-H$ colors turn to blue values, typically lying above and below the field sequence in the $K$ versus $J-H$ (Fig.~\ref{wise}) and the $J$ versus $J-H$ (Fig.~\ref{jhk2}) panels. This implies that the flux at 1.2 $\mu$m is notoriously suppressed and the flux at $\ge$2 $\mu$m  is enhanced.

A related behavior has been reported for L-type dwarfs in the field that are suspected to have young (intermediate) ages from the distinct low-gravity features in their optical and/or near-infrared spectra (e.g., \citealt{liu13,schneider14}; see also review by \citealt[and references therein]{luhman12}). For comparison purposes, in Fig.~\ref{wise} we plot three of these objects taken to the distance of the Pleiades: 2MASS\,J03552337$+$1133437 (L5, \citealt{reid06}), 2MASS\,J05012406$-$0010452 (L4, \citealt{reid08}), and PSO\,J318.5338$-$22.8603 (L7$\pm$1, \citealt{liu13}). Their 2MASS photometry was transformed into the UKIDSS system using the equations given in \citet{hodgkin09}, and spectral types are taken from the calibration of \citet{cruz09} and \citet{liu13}. All three are believed to be free-floating and have trigonometric parallaxes and {\sl WISE} photometry available in \citet{faherty13}, \citet{liu13}, and \citet{osorio14}. According to these authors, the three field sources may have an age most similar to that of the Pleiades, very likely within the interval 50--500 Myr. Their location in the color-magnitude diagrams nicely overlaps with the Pleiades. We conclude that our Pleiades member candidates may become benchmark objects for understanding the young population of isolated ultracool dwarfs  in the field.

Our data may also be useful in interpreting the properties of giant planets orbiting stars in the solar neighborhood. The four objects around HR\,8799 discovered by \citet{marois08,marois10}, and the planet companions to GJ\,504 \citep{kuzuhara13} and GU\,Psc \citep{naud14} are shown in the panels of Fig.~\ref{wise}. Although still debated, the age of the HR\,8799 system may be compatible with $\sim$100 Myr (see discussion in \citealt[and references therein]{oppenheimer13}). We derived the mean $J$, $H$, and $K_s$ photometry of all four HR\,8799\,bcde planets from the values published in the literature \citep{marois08,marois10,esposito13,oppenheimer13}; plotted error bars represent the (large) dispersion of the measurements. According to \citet{kuzuhara13}, the age of GJ\,504 is estimated at 160$^{+350}_{-60}$, in agreement with the Pleiades age. The revised photometry of the late-T/early-Y GJ\,504\,b companion provided by \citet{janson13} was employed. \citet{naud14} discussed at length that the GU\,Psc system, consisting of an M3 star and a T3.5 planet, is most likely a member of the AB Doradus moving group, which is coeval with the Pleiades \citep{luhman05}. To complete the list of late-L and T-type dwarfs from the literature with ages compatible with 120 Myr and known (or constrained) distance, we include in all panels of Fig.~\ref{wise} the T7 source CFBDSIR\,J214947.2$-$040308.9 (J2149$-$0403) discovered by \citet{delorme12}. According to the authors, this object is also very likely a member of the AB Dor moving group. All objects have been taken to the distance of the Pleiades in Fig.~\ref{wise}. In the case of the AB Doradus moving group we adopted an average distance of 45\,$\pm$\,10 pc, which explains the large uncertainties in the $J$ and $K$ magnitudes of AB Dor objects shown in the figure. The location of the three farthest companions (bcd) of HR\,8799 are compatible with the least-massive Pleiades population at the 1-$\sigma$ level. Only the innermost companion (e) displays a deviating $H-K_s$ color (although at the 2-$\sigma$ level). Interestingly, the location of GU\,Psc\,b in the diagrams appears to be quite similar to that of our faintest candidate (Calar~26), suggesting that Calar~26 may be a $\sim$T3.5 free-floating planet in the Pleiades. GJ\,504\,b and J2149$-$0403 fall at significantly fainter magnitudes; however, their locations in the color-magnitude panels of Fig.~\ref{wise} are compatible with a smooth extrapolation of the Pleiades sequence toward lower luminosities (considering the large error bars associated with the faintest Pleiades candidates). Our candidates, GU\,Psc\,b, J2149$-$0403, and GJ\,504\,b, might be tracing the substellar sequence from the star--brown dwarf frontier through 5--8 M$_{\rm Jup}$ at the age of 120 Myr.

As claimed by \citet{liu13}, PSO\,J318.5338$-$22.8603 is the reddest known field isolated dwarf, which is assigned L7$\pm$1 spectral type and effective temperature $T_{\rm eff} = 1210 ^{+40}_{-50}$ K. After comparison with our data, the reddest Pleiades member candidates with $J = 20.2-20.3$ mag may share a similar classification. Candidates extending down to fainter magnitudes are expected to have cooler effective temperatures and to display even later typings; they are probably entering into the methane-absorption regime (T types). This agrees with the location of the possible color turn-over observed in the Pleiades sequence and the tentative detection of methane absorption in the planets of HR\,8799 and GJ\,504 \citep{oppenheimer13,janson13}. Nevertheless, spectroscopic observations are strongly demanded to characterize the very low luminous domain of the Pleiades in more detail.

That our survey is sensitive to the detection of standard L--T transition and T-type dwarfs is demonstrated by the discovery of three sources with colors typical of these spectral classifications and proper motions in $\alpha$ greater than the defined astrometric selection of Fig.~\ref{pm} (see Sect.~\ref{astrometry}). These objects are included in Figs.~\ref{jhk2} and~\ref{wise}. Given their photometric and astrometric properties, they might be Pleiades member candidates since their proper motion error bars touch the circle of the astrometric selection criterion (less likely) or unrelated field late-L and T dwarfs, white dwarfs, and/or low-metallicity dwarfs.

\subsection{Comparison with evolutionary models \label{model}}
Two 120 Myr isochrones from the Lyon group are displayed in the top panels of Fig.~\ref{jhk2}: One track (dotted line) uses the magnitudes and colors predicted for each mass by model atmosphere synthesis, while the observable magnitudes of the other track (dashed line) were computed using bolometric corrections (see below). A solar metallicity was adopted for the two cases. Regarding the former track, we used the theoretical luminosities and temperatures given by \citet{chabrier00a}, and the colors computed with the BT-Settl model atmospheres by \citet{allard12}; we ensured that the computed colors are in the same photometric system as our data. The BT-Settl models account for the settling of some dust species from the photosphere, in contrast with the DUSTY models (which included dust formation, but once dust is formed, it remains in the atmosphere, \citealt{chabrier00a}), the COND models (which neglect the effect of dust opacity in the radiative transfer, \citealt{baraffe03}), and the NextGen models (which did not account for dust formation, \citealt{hauschildt99}). It is accepted that atmospheric dust formation \citep{tsuji96} becomes efficient at  temperatures typically below 2700 K and makes objects redder in all wavelengths \citep{allard97}. From the top panels of Fig.~\ref{jhk2}, it becomes apparent that the BT-Settl 120 Myr theoretical track does not reproduce the observations in two ways: the synthetic colors predict the appearance of methane absorption (sequence turnover) earlier than observed, and these colors are never as red as those of faint Pleiades member candidates.  

It is also customary to employ evolutionary models and empirical bolometric corrections (BCs) valid for high-gravity field dwarfs to convert predicted $T_{\rm eff}$ and luminosities into observables. This is because the Pleiades and field stars of related temperatures share similar properties. This would improve the comparison between the Pleiades sequence and the theory from $J = 15.5$ down to $J \sim 18.5$ mag; however, the reddening observed at fainter near-infrared magnitudes would still need to be fit. To overcome this problem, we built the 120 Myr isochrone shown with a dashed line in Fig.~\ref{jhk2} as follows: In the interval $J = 15.5-18.5$ mag we used the empirical BCs available in the literature for field dwarfs with a spectral classification between M6 and L3--L5 or 2700 K to 1800--1500 K \citep{dahn02,vrba04,golimowski04,stephens09}, and the predicted $T_{\rm eff}$'s and luminosities given by \citet{chabrier00a}. For later types, we obtained the average $J-H$ and $J-K_s$ colors of field dwarfs reported to have spectral features typical of low-gravity atmospheres and a reddish behavior at long wavelengths \citep[and references therein]{cruz09,gizis12,liu13}, and the BCs specifically derived for a few of them by \citet{todorov10}, \citet{osorio10}, and \citet{liu13}. These authors found differing BCs between the young and the standard field dwarfs of similar classification. This agrees with the statements made by \citet{luhman12} and \citet{faherty12} that low surface gravity L dwarfs require a new set of BC/absolute magnitude calibrations. BCs for young cooler types are lacking in the literature; the newly derived 120 Myr track (semi-empirical model) ends at $T_{\rm eff} \sim 1250$ K ($\sim$0.012 M$_\odot$). The uncertainty associated with this isochrone is $\pm$0.2 mag in the derived red near-infrared colors. As illustrated in Fig.~\ref{jhk2}, this isochrone nicely reproduces the trend delineated by the Pleiades sequence from the substellar limit through the deuterium-burning mass threshold: It is slightly more overluminous than the field sequence and becomes increasingly redder in $J-H$ and $J-K_s$ down to $J,K_s \approx 20.5, 18.0$ mag.

The NextGen, DUSTY, and COND solar-metallicity 120 Myr isochrones are depicted in the bottom panels of Fig.~\ref{jhk2}. We used the predicted observable magnitudes provided by the theory. None of these models is able to conveniently reproduce the Pleiades substellar sequence. Only the NextGen track successfully delineates the cluster trend at the brightest near-infrared magnitudes that include low-mass stars. This was previously discussed by, e.g., \citet{osorio97}, \citet{martin00}, and \citet{pinfield03}. Pleiades sources remain consistent with the NextGen model (and inconsistent with the DUSTY and COND models) down to $T_{\rm eff}$ values of $\sim$2400 K (or 0.055 M$_\odot$), coinciding with the expected temperature for dust formation. In the bottom panels of Fig.~\ref{jhk2} we show a combined NextGen$+$DUSTY$+$COND isochrone computed for solar metallicity and 120 Myr as follows: For temperatures between 2400 and 1100--1200 K, the DUSTY and COND predicted magnitudes were averaged to account for a situation intermediate between the extreme cases represented by the two models (i.e., formation of condensates, opacity increase due to the presence of dusty species, and partial sedimentation of grains). Below 1100 K, where all dust is supposed to lie below the photosphere, only the COND models were used. DUSTY and COND magnitudes were shifted to match the NextGen track that fits the data at the warmest temperatures. The resulting isochrone, which we remark was produced for illustrative purposes alone, is displayed with a continuous line in the bottom panels of Fig.~\ref{jhk2}. This track gives a better qualitative fit to the Pleiades sequence than the individual models, thus proving that dust formation and an increasing opacity at blue wavelengths provide a reasonable explanation for the observed reddening of the cluster sequence down to $J,K_s \approx 20.5, 18.0$ mag, $T_{\rm eff} \approx 1200$ K, and mass of $\approx$0.012 M$_\odot$. At fainter magnitudes or lower temperatures and masses, dust precipitation below the photosphere appears to gain importance.

  \begin{figure}
  \centering
  \includegraphics[width=0.48\textwidth]{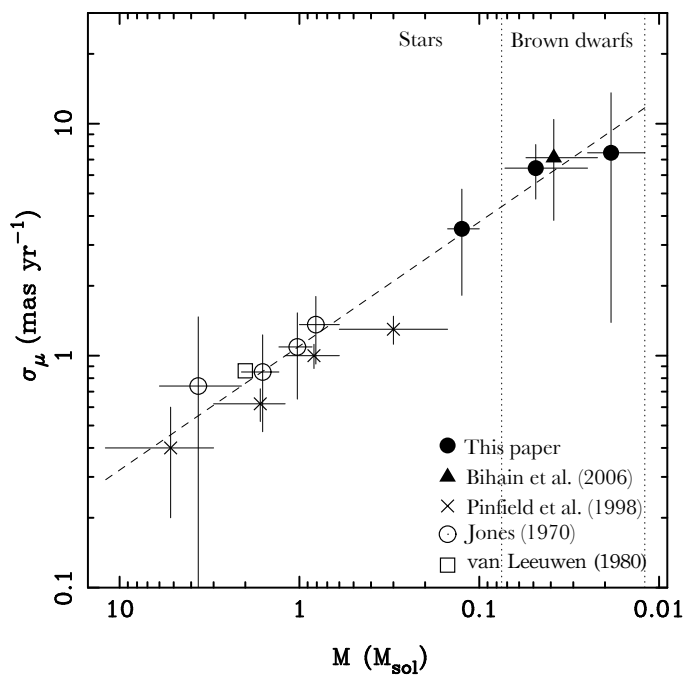}
  \caption{Pleiades proper motion dispersion plotted against mass. Data from various studies (including our own) covering from 12 to 0.012 M$_\odot$ are depicted with different symbols (see the legend within the diagram). Vertical error bars stand for the uncertainties in the motion-dispersion determination, and horizontal error bars account for mass intervals. The dashed line corresponds to the best fit to all data points excluding the least massive datum from \citet{pinfield98}. The slope of the fit is $-0.53$, suggesting near-equipartition of energy for stars and brown dwarfs in the cluster. Vertical dotted lines separate the stellar and the brown dwarf domains. \label{pmdispgraph}}
  \end{figure}

Note that the mass estimates (indicated in all panels of Figs.~\ref{jhk2} and~\ref{wise}) based on the newly derived isochrones differ from those obtained from the {BT-Settl} synthetical colors. In the luminosity interval of our study, the discrepancy can be as large as a factor of $\sim$2 in mass at the faintest magnitudes and for the $J$ band. Discrepancies are smaller and reversed for the $K_s$ filter. The masses provided by the NextGen$+$DUSTY$+$COND isochrone agree better with those obtained from the semi-empirical model. This latter model yields greater masses for a given observed magnitude. For example, the deuterium-burning mass limit would be located at $J, K_s \sim 20.9, 18.1$ mag (new 120 Myr isochrone), in contrast with the values of $J, K_s \sim 18.8, 18.6$ mag given by the BT-Settl synthetic colors track. Following the new derivation, Pleiades candidates Calar~21--26 have masses near the deuterium-burning
mass limit, thus becoming excellent targets for the observation of deuterium in their atmospheres.

\section{Proper motion dispersion \label{pmdisp}}
As shown in Tables~\ref{cands} and~\ref{cands2}, within the survey completeness, the proper motion standard deviations of individual Pleiades candidates typically fall between $\pm$1 and $\pm$3.8 mas\,yr$^{-1}$, the median value is $\pm$1.7 mas\,yr$^{-1}$ . This is generally smaller than the quoted astrometric uncertainties.  By assuming that  $\pm$1.7 mas\,yr$^{-1}$ is close to the true error in the proper motion determination of Pleiades candidates with magnitudes in the interval $J$ = 15.5--19, we derived that the proper motion dispersion (or scatter) of cluster members with masses between 0.072 and 0.025 M$_\odot$ (see Sect.~\ref{model}) is $\sigma_\mu$\,=\,6.7\,$\pm$\,1.7 mas\,yr$^{-1}$. The averaged motions are estimated at $\mu_\alpha\,{\rm cos}\,\delta$ = 19.6 mas\,yr$^{-1}$ and $\mu_\delta$ = $-$41.9 mas\,yr$^{-1}$. For fainter magnitudes and smaller masses ($J$\,=\,19--21 mag, 0.012--0.025 M$_\odot$), proper motion measurements have larger uncertainties and standard deviations, yet we obtained a motion dispersion of $\sigma_\mu$\,=\,9.7\,$\pm$\,6.1 mas\,yr$^{-1}$ and mean motions of $\mu_\alpha\,{\rm cos}\,\delta$ = 21.6 mas\,yr$^{-1}$ and $\mu_\delta$ = $-$47.6 mas\,yr$^{-1}$.  The mean motions of the two mass intervals agree with that of the cluster \citep{lotkin03} at the 1-$\sigma$ level. The proper motion dispersion estimates, $\sigma_\mu$, include the scatter of the astrometric errors. In what follows it is assumed that the distribution of proper motions and errors is Gaussian. The results can be corrected for the errors by applying the equation\begin{equation}
\sigma^{0}_{\mu} = \sqrt{\sigma^2_\mu - {\rm err}^2_{\sigma_\mu}}
.\end{equation}
The corrected proper motion dispersions, $\sigma^{0}_{\mu}$, are 6.4\,$\pm$\,1.7 mas\,yr$^{-1}$ and 7.5\,$\pm$\,6.1 mas\,yr$^{-1}$ for the mass intervals 0.072--0.025 M$_\odot$ and 0.025--0.012 M$_\odot$. With our data, the detection of motion dispersion in the Pleiades brown dwarf regime is at the level of $\sim$3.5 $\sigma$ for cluster objects with 0.072--0.025 M$_\odot$. 

The corrected values are plotted against mass in Fig.~\ref{pmdispgraph}. These data are complemented with the proper motion dispersion obtained for Pleiades very low-mass stars with masses ranging from 0.1 to 0.15 M$_\odot$ identified in the $JH$ survey. These stars have $J = 14.5-15.05$ mag; they were selected following the same astrometric and photometric criteria as described in Sect.~\ref{astrometry}. Additional proper motion dispersions obtained for different mass intervals by other groups are also plotted in Fig.~\ref{pmdispgraph}: stars from \citet{jones70}, \citet{vanleeuwen80}, and \citet{pinfield98}, and brown dwarfs (0.022--0.055 M$_\odot$) from \citet{bihain06}. Our determinations for substellar cluster members and that of \citet{bihain06} agree within the quoted uncertainties as illustrated in the figure. Substellar Pleiades objects show a proper motion dispersion higher
by a factor of $\sim$6--7 than solar-mass
stars. By excluding the least-massive bin of \citet{pinfield98}, which appears to deviate from a straight line in the log-log representation of Fig.~\ref{pmdispgraph}, we fit all data points following a power-law expression: $\sigma^0_\mu \sim m^{\gamma}$. We found $\gamma$\,=\,$-0.53$ and a fit {\sl rms} of $\pm$0.074 in units of log\,($\sigma^0_\mu$). The resulting fit is shown in Fig.~\ref{pmdispgraph} by the dashed line. The measured slope is fully compatible with the energy equipartition theorem, where the velocity dispersion scales with mass as $\sigma^0_\mu \sim m^{-0.5}$, which agrees with the presence of mass segregation within the mass coverage shown in Fig.~\ref{pmdispgraph} and with the dynamical relaxation status of the cluster as discussed in \citet[and references therein]{jameson02}. At the distance of 133.5 pc \citep{soderblom05}, the measured substellar proper motion dispersions translate into velocity dispersions of $\pm$4--5\,$\pm$\,1.0 km\,s$^{-1}$.

   \begin{figure*}
   \centering
   \includegraphics[width=0.48\textwidth]{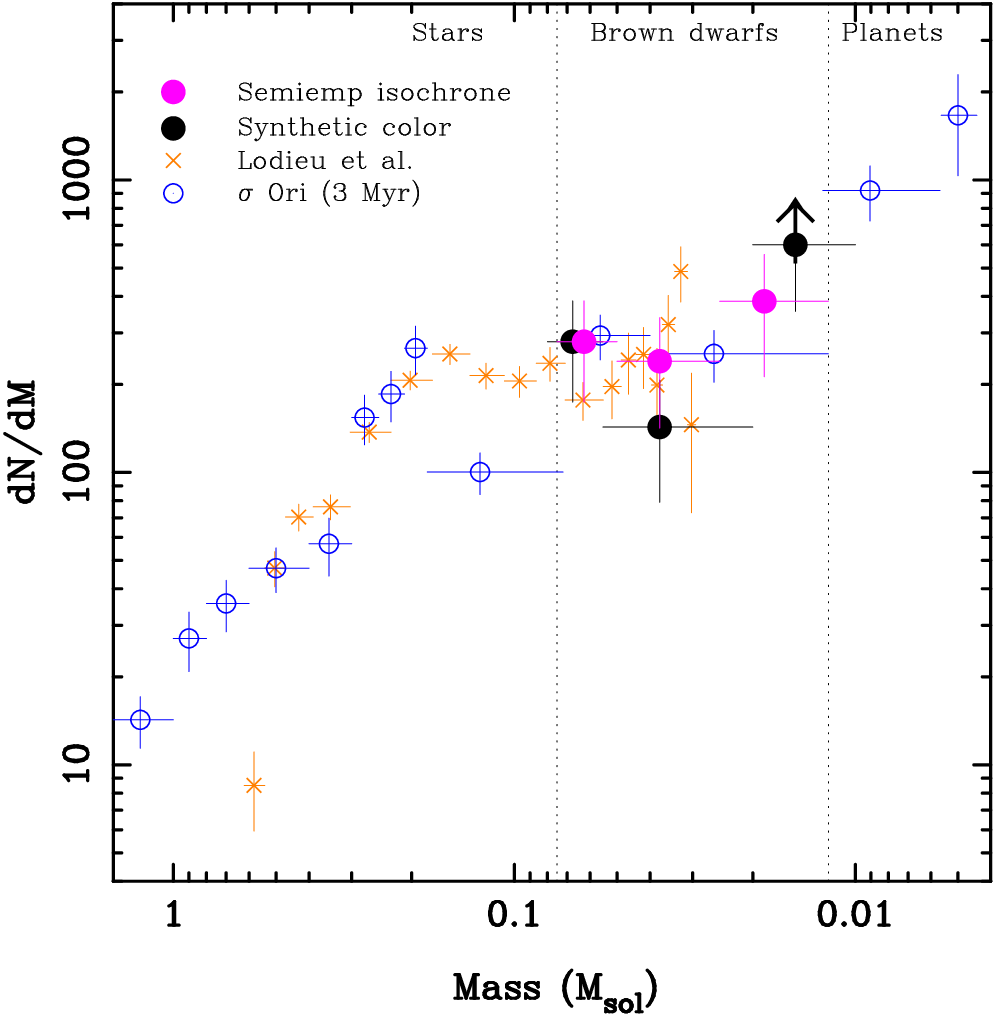} ~~~~~
   \includegraphics[width=0.48\textwidth]{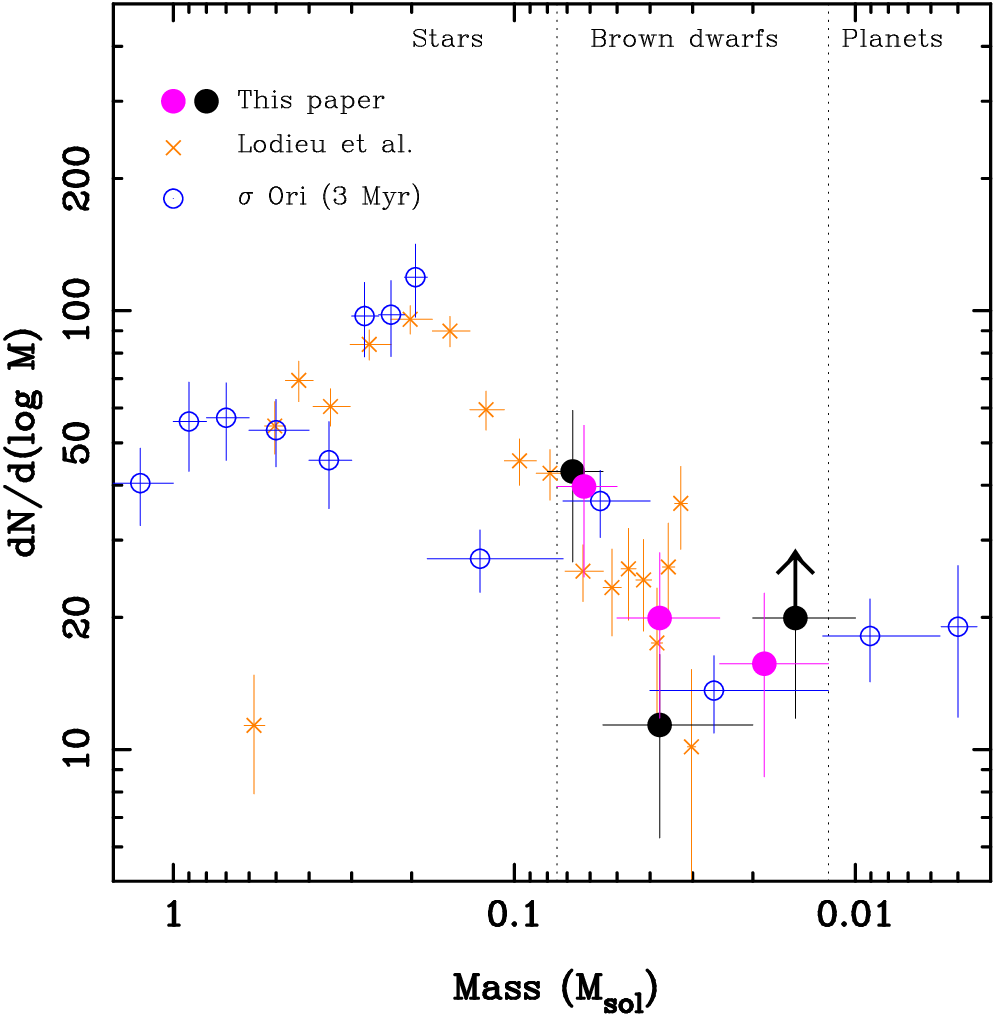}
   \caption{Pleiades mass spectrum (left) and mass function (right) obtained by combining our data (black and magenta filled circles) with data from \citet[brown crosses]{lodieu12}. Also included is the mass spectrum and mass function of the $\sigma$~Orionis cluster ($\sim$3 Myr) from \citet[blue open circles]{pena12}. All functions are normalized to the total number of brown dwarfs in the mass interval 0.075--0.03 M$_\odot$ found in our survey. The agreement between the two clusters is remarkable. Vertical dotted lines stand for the hydrogen and deuterium-burning mass limits at $\approx$0.072 and $\approx$0.012 M$_\odot$. Vertical error bars indicate Poisson uncertainties, and horizontal error bars correspond to mass intervals. The incomplete mass bin is marked with an upper arrow. Both axes are in logarithmic scale. }
              \label{mf}
    \end{figure*}

\section{Pleiades substellar mass function \label{massf}}
To build the cluster mass spectrum of the surveyed $\sim$0.8 deg$^2$ area (number of objects per linear mass unit, d$N$/d$m$), we first counted the number of Pleiades member candidates per $J$ and $K_s$ magnitude interval (similar to the luminosity function) and then applied the mass--magnitude relationship given by the evolutionary models. The solar-metallicity 120 Myr isochrones discussed in the previous section were considered: The isochrone of synthetic colors spanning a large mass interval and covering the magnitude range of our survey, and the semi-empirical isochrone extending from the low-mass star domain down to the deuterium-burning mass boundary at $\approx$0.012 M$_\odot$. For the former isochrone we used the $K_s$-band predictions because these are closer to those of the semi-empirical track. For the latter track, either $J$ or $K_s$ magnitudes yield similar results. Given the small number of Pleiades candidates (19, see Section~\ref{seq}), we fixed the minimum number of objects per mass bin at $\ge$5. This is to give statistical weight to the mass spectrum derivation, which is depicted in the left panel of Fig.~\ref{mf}. The two substellar-mass spectra are consistent within the quoted error bars. The incomplete mass bin is indicated with an upper arrow in the figure.

The Pleiades mass spectrum appears rather flat within the uncertainties from the substellar frontier to the deuterium-burning mass limit. Our data points in Fig.~\ref{mf} are consistent with the power law expression d$N$/d$m$ $\sim$ $m^{-\alpha}$, where $\alpha$ lies in the interval 0--1. This functional form and index range are widely discussed in other studies dealing with young star clusters and star associations with ages between 1 and 100 Myr (e.g., \citealt[and references therein]{moraux03,bihain06,lodieu12,scholz12,luhman12}). There is no obvious mass cutoff in the Pleiades cluster down to $\approx$0.012 M$_\odot$, which roughly coincides with the completeness of the main $JH$ survey. Furthermore, our observations suggest that the cluster mass function extends beyond the deuterium-burning
mass borderline into the planetary-mass regime. 

For comparison purposes, we included in the left panel of Fig.~\ref{mf} the Pleiades (0.60--0.03 M$_\odot$) and $\sigma$~Orionis (1--0.004 M$_\odot$) mass spectra obtained by \citet{lodieu12} and \citet{pena12}. We chose the $\sigma$~Orionis cluster because its mass function was derived for a wide range of masses within the substellar domain.  Also shown in the right panel of Fig.~\ref{mf} are the two clusters mass functions that consider the number of objects per logarithmic mass unit [d$N$/dlog($m$)] as defined by \citet{salpeter55}. All functions are normalized to the total number of Pleiades candidates in the $JH$ main survey for the common mass range 0.03--0.075 M$_\odot$. Interestingly, they have been derived using the prescriptions described before and the same solar-metallicity evolutionary models. Therefore, the mass functions can be safely compared. 

The Pleiades is $\sim$40 times older than the $\sigma$~Orionis cluster ($\sim$3 Myr, \citealt{osorio02,oliveira04}), yet the two regions display fairly similar mass spectrum and mass function for all masses in common, from $\sim$0.5 through $\sim$0.012 M$_\odot$. In the mass spectrum form, the number of objects between solar-mass stars and $\sim$0.2 M$_\odot$ increases toward low masses in both clusters. Between $\sim$0.2 M$_\odot$ and $\sim$0.03 M$_\odot$ in the substellar domain the mass spectrum flattens and apparently changes to a rising slope for smaller masses into the planetary domain. The similar behavior between the Pleiades and $\sigma$~Orionis mass functions suggests that even though the Pleiades is believed to be dynamically evolved (Sect.~\ref{pmdisp}), it has not lost (or evaporated) a significant amount of its low-mass content (brown dwarf and massive planet;  see also \citealt{moraux04,bihain06}), or that if any, the relative object loss between susbtellar members and low-mass stars occurs in a related manner in both clusters (assuming that the evolutionary models are correct for these two very different ages). 

We caution that the Pleiades mass function and the discussion presented here are based on a rather small explored region, which approximately covers $\sim$3\%~of the total cluster field. This study would benefit from a similar or even deeper survey that extends to a significant fraction of the Pleiades area. Additionally, follow-up observations are needed to confirm the cluster membership of our candidates, particularly the faintest ones, to construct a more robust Pleiades mass function in the planetary domain. The extrapolation of the Pleiades mass function illustrated in Fig.~\ref{mf} into very small masses using a power law with indices of 0.2, 0.6, and 1.0 yields that 150--850 planetary-mass sources between 0.012 and 0.001 M$_\odot$ may be free-floating in the cluster. The least-massive ones would have effective temperatures like those of Earth and luminosities three orders of magnitude lower than the candidates of this work. Detecting them is indeed challenging even for 10 m class telescopes; they are targets for next-generation telescopes.

\section{Conclusions}

We carried out an astrometric and photometric search in the Pleiades covering an area of $\sim$0.8 deg$^2$ around the cluster center and using $J$- and $H$-band images taken 9 yr apart with the near-infrared cameras of the Calar Alto 3.5 m telescope. The exploration was sensitive to Pleiades member candidates spanning the magnitude interval $J$ = 15.5--21.5, this is from the substellar limit at the age (120 Myr) and distance (133.5 pc) of the cluster through 0.008--0.010 M$_\odot$ according to the evolutionary models by \citet{chabrier00a}. The completeness of the survey lies at $J \approx 20-20.5$ mag and $\approx$0.012 M$_\odot$, that is, the deuterium-burning mass threshold. This is about 1--2 mag fainter than previous works \citep{bihain06,lodieu12}. The typical $JH$ astrometric precision is about $\pm$6 mas\,yr$^{-1}$; third- (and fourth-) epoch UKIDSS $K$-band images and images acquired by us with HAWK-I on the VLT and LIRIS on the WHT separated by 11--15 yr in time allowed us to reduce the proper motion error bar to $\pm$1--4  mas\,yr$^{-1}$ for some selected candidates. 

Pleiades member candidates were picked up with proper motions compatible with the one distinctive of the cluster within a radius of $\pm$13 mas\,yr$^{-1}$, and with magnitudes and colors following the known cluster sequence in the interval $J$ = 15.5--18.8 mag, and $Z_{\rm{UKIDSS}} - J \ge 2.3$ mag or no $Z$-band detection for $J > 18.8$ mag. We found that about 50\%~of the astrometric candidates fail the photometric criteria, which left us with 19 Pleiades candidates that complied with all the required conditions for being true cluster members. Most of them have $W1$ and $W2$ photometry with S/N $\ge$ 3 available from the {\sl WISE} catalog. The Pleiades sequence from $J$ = 15.5 to 20.3 mag, or $K_s$ = 14.5 to 18 mag, became increasingly red in all studied colors; furthermore, the colors of candidates at $J = 20.2-20.3$ mag are significantly redder by 0.5--1 mag than those of high-gravity field L-type dwarfs. This agrees with the photometric properties reported for young L dwarfs in the field (e.g., \citealt{cruz09,osorio10,faherty12,liu13,schneider14}). This signature is not well reproduced by the 120 Myr theoretical isochrone \citep{chabrier00a} with the synthetic colors computed from the BT-Settl model atmospheres \citep{allard12}. This isochrone predicts that the near-infrared colors of Pleiades substellar members turn to blue indices at about 1 mag brighter, in clear contrast with the observations. A semi-empirical 120-Myr model built from the theoretical luminosities and $T_{\rm eff}$'s by \citet{chabrier00a} and the bolometric corrections available in the literature for field young L dwarfs nicely reproduces the Pleiades photometric sequence from $J, K_s$ = 15.5, 14.5 to $\approx$20.5, 18.0 mag.

For fainter $J$ and $K_s$ magnitudes, or masses below $\approx$0.012 M$_\odot$ at the age and distance of the cluster, our Pleiades candidates appear to turn slightly to blue $J-H$ and $J-K_s$ colors, while the  few {\sl WISE} data available to us suggest that the red behavior of the $K_s-W2$ index continues to rise. This might be due to an increasing object contamination at the faintest end of our survey or to the appearance of methane absorption in the atmospheres of Pleiades members. The location of the three farthest giant planets and the T3.5 type planet of the $\sim$100
Myr systems HR\,8799 and GU\,Psc \citep{marois08,marois10,naud14} in color-magnitude diagrams was reasonably explained by the disposition and colors of our least-luminous candidates. This would support the planetary mass of the companion objects of HR\,8799 and GU\,Psc. Follow-up spectroscopic observations are necessary for a proper characterization of the Pleiades candidates.

The proper motion dispersion of Pleiades brown dwarfs (corrected for the errors of the proper motion measurements) was determined at 6.4\,$\pm$\,1.7 mas\,yr$^{-1}$ and 7.5\,$\pm$\,6.1 mas\,yr$^{-1}$ for the mass intervals 0.072--0.025 and 0.025--0.012 M$_\odot$. This is a factor of $\sim$6--7 times higher than the velocity dispersion observed for solar-mass stars in the cluster \citep{jones70,pinfield98}, which agrees with the energy equipartition where the velocity dispersion scales with $m^{-0.5}$. This indicates that brown dwarfs are dynamically relaxed at the Pleiades age. However, from our computations of the Pleiades substellar mass function combined with the stellar mass function obtained by \citet{lodieu12}, it seems that the cluster has expanded, but has not lost a significant amount of its brown dwarf and massive planetary-mass object content relative to the low-mass stellar content. We found  that the Pleiades mass function probably softly increases toward the planetary-mass domain, which was previously reported for the 40-times younger $\sigma$~Orionis cluster by \citet{pena12}. The Pleiades and $\sigma$~Orionis mass functions compare well from $\sim$0.5 through $\sim$0.012 M$_\odot$, which supports the universality of the mass function. According to evolutionary models, the new low-mass discoveries in the Pleiades cluster may have masses of about 0.012 M$_\odot$, effective temperatures ranging from 800 to 1300 K, and intermediate surface gravities around log\,$g$ = 4.0 (cm\,s$^{-2}$). They may become reference objects for interpreting the observations of the solar neighborhood ultracool population and giant planets that orbit stars.

\begin{acknowledgements}
We thank E.~L.~Mart\'\i n for useful discussions related to the topic of this article, we also thank H$.$ Bouy for discussions on proper motion measurements. We thank the anonymous referee for comments and suggestions that have improved this paper. Based on observations collected at the Centro Astron\'omico Hispano Alem\'an (CAHA) at Calar Alto, operated jointly by the Max-Planck Institut f\"ur Astronomie and the Instituto de Astrof\'\i sica de Andaluc\'\i a (CSIC). Also based on observations made with the William Herschel Telescope (WHT) operated on the island of La Palma by the Isaac Newton Group in the Spanish Observatorio del Roque de los Muchachos of the Instituto de Astrof\'\i sica de Canarias. Also based on observations collected at the European Organisation for Astronomical Research in the Southern Hemisphere, Chile (proposal 088.C-0328). This work is based in part on data obtained as part of the UKIRT Infrared Deep Sky Survey. This research has made use of the SIMBAD database, operated at CDS, Strasbourg, France. This work was financed by the Spanish Ministry of Economy and Competitiveness through the project AYA2011-30147-C03-03, and by Sonderforschungsbereich SFB 881 "The Milky Way System" (subprogram B6) of the German Research Foundation. M$.$ C$.$ G$.$ O$.$ acknowledges the support of a JAE-Doc CSIC fellowship co-funded with the European Social Fund under the program {\em Junta para la Ampliaci\'on de Estudios}.

\end{acknowledgements}

\bibliographystyle{aa} 
\bibliography{2.bib}


\addtocounter{table}{-2}
\begin{landscape}
\begin{table*}
\caption{Proper motion Pleiades candidate selection. \label{cands}}
\scriptsize
\begin{tabular}{lcccccccccl}
\hline\hline
\multicolumn{11}{l}{Two or more proper motion measurements and photometry ($J = 15.5-19.0$) consistent with cluster sequence, unresolved candidates.} \\
\hline
Object &  RA, DEC (J2000) & $\mu_\alpha$ cos $\delta$, $\mu_\delta$ & $N$\tablefootmark{a} & $J$\tablefootmark{b} & $H$\tablefootmark{b} & $K_s$\tablefootmark{c} & $W1$ & $W2$ & $\sigma (\mu_\alpha, \mu_\delta)$\tablefootmark{d} & Other name \\
       & ($^{\rm h}$ $^{\rm m}$ $^{\rm s}$),  ($^{\circ}$ $'$ $''$) & (mas\,yr$^{-1}$) &  & (mag) & (mag) & (mag) & (mag) & (mag) & (mas\,yr$^{-1}$) &  \\
\hline
Calar Pleiades  8 &  3 43 34.19 $+$25 02 16.6 &  21.2 $\pm$  5.6,  $-$40.1 $\pm$  6.5 &  5 & 15.56 $\pm$ 0.04 & 14.88 $\pm$ 0.01 & 14.57 $\pm$ 0.01 & 14.36 $\pm$ 0.03 & 14.14 $\pm$ 0.05 &  1.6, 3.8 &                           \\ 
Calar Pleiades  9 &  3 42 03.09 $+$24 03 41.1 &  25.5 $\pm$  5.6,  $-$33.9 $\pm$  6.5 &  2 & 15.59 $\pm$ 0.02 & 15.00 $\pm$ 0.01 & 14.67 $\pm$ 0.01 & 14.55 $\pm$ 0.03 & 14.40 $\pm$ 0.05 &           &                           \\ 
Calar Pleiades 10 &  3 46 14.06 $+$23 21 56.4 &  26.5 $\pm$  5.7,  $-$41.2 $\pm$  6.6 &  1 & 15.67 $\pm$ 0.02 & 15.02 $\pm$ 0.02 & 14.62 $\pm$ 0.01 & 14.46 $\pm$ 0.03 & 14.31 $\pm$ 0.05 &           &  UGCS J034614.06$+$232156.4 \\ 
Calar Pleiades 11\tablefootmark{e} &  3 42 18.02 $+$24 55 10.0 &  12.5 $\pm$  5.6,  $-$36.8 $\pm$  6.5 &  1 & 15.87 $\pm$ 0.04 & 15.28 $\pm$ 0.01 & 14.85 $\pm$ 0.01 & 14.61 $\pm$ 0.03 & 14.42 $\pm$ 0.06 &           &  UGCS J034218.01$+$245509.9 \\ 
Calar Pleiades 12 &  3 42 48.18 $+$24 04 01.2 &  13.2 $\pm$  5.7,  $-$45.8 $\pm$  6.6 &  3 & 16.05 $\pm$ 0.04 & 15.42 $\pm$ 0.01 & 14.94 $\pm$ 0.01 & 14.76 $\pm$ 0.03 & 14.48 $\pm$ 0.06 &  2.0, 1.0 &  Cl* Melotte 22 BPL 45    \\ 
Calar Pleiades 13 &  3 41 43.94 $+$25 03 30.5 &  16.8 $\pm$  5.7,  $-$49.6 $\pm$  6.6 &  1 & 16.37 $\pm$ 0.04 & 15.68 $\pm$ 0.02 & 15.13 $\pm$ 0.01 & 14.87 $\pm$ 0.03 & 14.53 $\pm$ 0.06 &           &                           \\ 
Calar Pleiades 14 &  3 45 50.63 $+$23 44 36.8 &  17.0 $\pm$  5.6,  $-$48.3 $\pm$  6.6 &  2 & 16.39 $\pm$ 0.04 & 15.78 $\pm$ 0.01 & 15.26 $\pm$ 0.01 & 14.90 $\pm$ 0.04 & 14.78 $\pm$ 0.09 &           &  Cl* Melotte 22 NPNPL 1   \\ 
Calar Pleiades 15 &  3 42 30.59 $+$25 02 39.3 &  18.8 $\pm$  5.8,  $-$41.4 $\pm$  6.7 &  2 & 17.01 $\pm$ 0.03 & 16.34 $\pm$ 0.01 & 15.75 $\pm$ 0.02 & 15.15 $\pm$ 0.03 & 15.12 $\pm$ 0.09 &           &  UGCS J034230.58$+$250239.2 \\ 
Calar Pleiades 16 &  3 47 46.78 $+$25 35 16.6 &  17.8 $\pm$  5.6,  $-$45.0 $\pm$  6.7 &  2 & 17.54 $\pm$ 0.04 & 16.76 $\pm$ 0.02 & 16.15 $\pm$ 0.03 &                  &                   &           &  UGCS J034746.77$+$253516.5 \\ 
Calar Pleiades 17 &  3 45 11.73 $+$23 41 43.6 &  28.7 $\pm$  5.9,  $-$38.7 $\pm$  6.8 &  1 & 17.65 $\pm$ 0.04 & 16.83 $\pm$ 0.02 & 16.12 $\pm$ 0.02 & 15.80 $\pm$ 0.05 & 15.81 $\pm$ 0.17 &           &  Cl* Melotte 22 NPNPL 3   \\ 
Calar Pleiades 18 &  3 45 33.30 $+$23 34 34.3 &  21.6 $\pm$  6.0,  $-$41.4 $\pm$  7.0 &  3 & 18.18 $\pm$ 0.03 & 17.24 $\pm$ 0.02 & 16.50 $\pm$ 0.03 & 16.12 $\pm$ 0.06 & 15.82 $\pm$ 0.15 &  1.5, 1.0 &  UGCS J034533.30$+$233434.2 \\ 
Calar Pleiades 19 &  3 48 15.65 $+$25 50 08.8 &  16.1 $\pm$  6.3,  $-$42.4 $\pm$  7.2 &  3 & 18.49 $\pm$ 0.03 & 17.52 $\pm$ 0.02 & 16.73 $\pm$ 0.05 & 16.23 $\pm$ 0.07 & 16.57 $\pm$ 0.31 &  1.7, 1.5 &  UGCS J034815.64$+$255008.9 \\ 
Calar Pleiades 20 &  3 45 58.48 $+$23 41 53.8 &  19.4 $\pm$  6.4,  $-$40.8 $\pm$  7.2 &  2 & 18.62 $\pm$ 0.03 & 17.48 $\pm$ 0.01 & 16.74 $\pm$ 0.04 & 15.75 $\pm$ 0.05 & 15.49 $\pm$ 0.12 &           &  Cl* Melotte 22 NPNPL 4   \\ 
Calar Pleiades 21 &  3 47 00.36 $+$23 21 56.8 &  25.4 $\pm$  9.5,  $-$37.2 $\pm$  9.5 &  3 & 20.23 $\pm$ 0.08 & 18.74 $\pm$ 0.03 & 17.73 $\pm$ 0.03\tablefootmark{f} & 16.48 $\pm$ 0.08 & 16.31 $\pm$ 0.25 &  2.7, 2.7 &                           \\ 
Calar Pleiades 22 &  3 46 00.39 $+$23 19 33.9 &  18.4 $\pm$  9.1,  $-$57.6 $\pm$ 10.3 &  3 & 20.29 $\pm$ 0.07 & 18.92 $\pm$ 0.02 & 17.83 $\pm$ 0.13 &                &                  &  1.4, 7.3 &                             \\ 
Calar Pleiades 23\tablefootmark{e} &  3 42 55.13 $+$24 09 02.6 &  31.6 $\pm$ 16.0,  $-$54.0 $\pm$ 16.3 &  2 & 20.60 $\pm$ 0.22 & 19.69 $\pm$ 0.11 & 18.60 $\pm$ 0.18 & 16.99 $\pm$ 0.13 & 16.80 $\pm$ 0.40 &           &                           \\ 
Calar Pleiades 24 &  3 46 48.67 $+$23 28 48.7 &  14.6 $\pm$  9.1,  $-$45.5 $\pm$ 15.8 &  2 & 20.65 $\pm$ 0.13 & 20.04 $\pm$ 0.05 & 18.83 $\pm$ 0.33 &                  &                   &           &                           \\ 
Calar Pleiades 25 &  3 48 26.62 $+$22 51 53.9 &  19.8 $\pm$ 10.3,  $-$49.4 $\pm$ 12.5 &  3 & 20.83 $\pm$ 0.15 & 19.74 $\pm$ 0.07 & 18.46 $\pm$ 0.21 & 17.43 $\pm$ 0.16 &                   & 10.3,12.5 &                           \\ 
Calar Pleiades 26 &  3 45 16.09 $+$23 35 07.3 &  20.2 $\pm$  8.1,   $-$42.0 $\pm$  9.2 &  2 & 21.15 $\pm$ 0.14 & 20.61 $\pm$ 0.12 & 19.40 $\pm$ 0.50\tablefootmark{g} & 16.15 $\pm$ 0.06\tablefootmark{h} & 15.58 $\pm$ 0.13\tablefootmark{h} &           &                           \\ 
\hline
\hline
\end{tabular}
\begin{tabular}{lccccccc}
\multicolumn{8}{l}{One proper motion measurement and $J \ge 20.0$ mag, unresolved candidates.} \\
\hline
Object &  RA, DEC (J2000) & $\mu_\alpha$ cos $\delta$, $\mu_\delta$ & $J$\tablefootmark{b} & $H$\tablefootmark{b} & $K_s$\tablefootmark{c} & $W1$ & $W2$ \\
       & ($^{\rm h}$ $^{\rm m}$ $^{\rm s}$),  ($^{\circ}$ $'$ $''$) & (mas\,yr$^{-1}$) & (mag) & (mag) & (mag) & (mag) & (mag)  \\
\hline
Calar Pleiades 27\tablefootmark{e} &  3 46 12.13  $+$23 18 15.7  &   26.4 $\pm$  8.6,  $-$45.2 $\pm$ 11.3 &    20.01 $\pm$ 0.15 & 20.61 $\pm$ 0.07 &  $>$18.50         & & \\ 
Calar Pleiades 28 &  3 46 37.89  $+$23 25 27.3  &   21.4 $\pm$  8.0,  $-$43.6 $\pm$ 15.5 &    20.52 $\pm$ 0.12 & 20.21 $\pm$ 0.05 &  $>$18.48         & & \\ 
Calar Pleiades 29 &  3 42 54.18  $+$24 08 11.5  &   27.4 $\pm$  9.4,  $-$36.6 $\pm$  9.6 &    20.80 $\pm$ 0.13 & 20.65 $\pm$ 0.20 &  $>$18.65         & & \\ 
Calar Pleiades 30 &  3 48 01.49  $+$25 36 49.2  &   27.6 $\pm$ 13.0,  $-$49.9 $\pm$ 15.5 &    21.19 $\pm$ 0.11 & 20.39 $\pm$ 0.05 &  $>$18.64         & & \\ 
Calar Pleiades 31 &  3 45 42.49  $+$23 43 49.4  &   17.9 $\pm$ 13.0,  $-$41.4 $\pm$ 15.5 &    21.37 $\pm$ 0.20 & 20.82 $\pm$ 0.09 &  18.99 $\pm$ 0.50 & & \\ 
\hline
\hline
\end{tabular}
\begin{tabular}{lccccccc}
\multicolumn{8}{l}{Resolved candidates and $J \ge 20$ mag.} \\
\hline
Object &  RA, DEC (J2000) & $\mu_\alpha$ cos $\delta$, $\mu_\delta$ & $J$\tablefootmark{b} & $H$\tablefootmark{b} & $K_s$\tablefootmark{c} & $W1$ & $W2$ \\
       & ($^{\rm h}$ $^{\rm m}$ $^{\rm s}$),  ($^{\circ}$ $'$ $''$) & (mas\,yr$^{-1}$) & (mag) & (mag) & (mag) & (mag) & (mag)  \\
\hline
R1 & 3 43 00.72  $+$25 05 04.6 &   21.5 $\pm$ 10.0,  $-$34.0 $\pm$ 11.0   &   19.97 $\pm$ 0.08 & 18.97 $\pm$ 0.05  &  18.47 $\pm$ 0.14  &  17.46 $\pm$ 0.20 &                  \\ 
R2 & 3 48 01.30  $+$23 08 13.1 &   15.6 $\pm$ 10.0,  $-$54.6 $\pm$ 11.0   &   19.98 $\pm$ 0.06 & 18.68 $\pm$ 0.14  &  18.24 $\pm$ 0.13  &  16.86 $\pm$ 0.16 & 17.07 $\pm$ 0.47 \\ 
R3 & 3 48 28.18  $+$23 12 33.0 &   26.2 $\pm$ 11.0,  $-$35.4 $\pm$ 12.0   &   20.41 $\pm$ 0.10 & 19.11 $\pm$ 0.16  &  $>$18.69          &                   &                  \\ 
R4 & 3 49 49.61  $+$22 53 55.5 &   17.7 $\pm$ 12.0,  $-$38.1 $\pm$ 13.0   &   20.76 $\pm$ 0.09 & 20.20 $\pm$ 0.10  &  $>$18.51          &                   &                  \\ 
R5 & 3 48 18.32  $+$25 27 22.7 &   16.2 $\pm$ 13.0,  $-$35.6 $\pm$ 15.5   &   21.18 $\pm$ 0.15 & 20.51 $\pm$ 0.09  &  $>$18.63          &                   &                  \\ 
\hline
\end{tabular}
\\
\tablefoottext{a}{Number of proper motion measurements. The number of observing epochs has to be increased by 1.}
\tablefoottext{b}{This paper.}
\tablefoottext{c}{From UKIDSS GCS, with a few indicated exceptions.}
\tablefoottext{d}{Standard deviation of the mean proper motion measurement computed for $N \ge 3$.} 
\tablefoottext{e}{Close to $J$- and/or $H$-band frame border, large uncertainty in both photometric and astrometric data.}
\tablefoottext{f}{LIRIS $K_s$ measurement.}
\tablefoottext{g}{UKIDSS GCS $K_s$ measurement obtained by us (see Sect.~\ref{observations}).}
\tablefoottext{h}{The {\sl WISE} source is located at 1\farcs99 from the $H$-band object (see text).}
\end{table*}
\end{landscape}

\end{document}